# Directly Mapping Interacting Components to Complex Systems' Emergent Properties

## Lina Yan[1]†, Jeffrey Huy Khong[2]†, Aleksandar Kostadinov[3], Wen-Jun Chen[4], Jerry Ying Hsi Fuh[1], Chih-Ming Ho[1,5]*

**Affiliations:**

[1]Department of Mechanical Engineering, National University of Singapore; Singapore, 119077, Singapore.

[2]Department of Psychology, University of California, Los Angeles; Los Angeles, 90095, United States of America.

[3]Department of Biological Sciences, National University of Singapore; Singapore, 119077, Singapore.

[4]Department of Power Engineering, National Tsinghua University; Hsinchu City, 30013 Taiwan.

[5]Department of Mechanical and Aerospace Engineering, University of California, Los Angeles; Los Angeles, 90095, United States of America.

*Corresponding author. Email: chihming@g.ucla.edu

†These authors contributed equally to this work.

**Abstract:**

Emergent behavior in complex systems arises from nonlinear interactions among components, yet the intricate nature of self-organization often obscures the underlying causal relationships, long regarded as the "holy grail" of complexity research. To address this challenge, we adopted an inductive, mechanism-agnostic approach to characterize how diseased biological systems respond to therapeutic interventions, leading to the discovery of the Complex System Response (CSR) equation, a deterministic formulation that quantitatively connects component interactions with emergent behaviors, validated across ~30 disease models. In this study, the main goal is to extend the CSR framework beyond biology to validate its applicability in engineering systems and urban social dynamics. Our results reveal that the CSR equation embodies the same systemic principles governing physical, chemical, biological, and social complex systems.

**Main Text:**

The science of complex systems (1, 2) constitutes a transformative field of physics that seeks to understand phenomena from a holistic, system-level perspective. Complex systems are pervasive in nature and society, encompassing the human body, engineering systems, and social structures. A complex system is typically defined by the following four attributes (3, 4):

1. It comprises numerous interacting components.
2. These components undergo self-organization, whereby their interactions propagate and evolve into a nonlinear, dynamic system.
3. The intricacy of self-organization makes it extremely challenging to infer emergent properties directly from the underlying component interactions.

4. The system exhibits adaptivity and robustness.

Contemporary scientific inquiry largely follows the reductionist paradigm, wherein systems are decomposed into subsystems and analyzed in parts (4). At fine scales, stochastic components exchange information across networks (5, 6), and their local interactions give rise to self-organized, large-scale patterns that drive the system toward a critical state (7, 8). This self-organization may proceed through multiple hierarchical stages (9,10) until coherent macroscopic properties emerge. However, as nonlinear self-organizations accumulate and critical transitions occur abruptly, it becomes insurmountable to connect the emergent properties back to their microscopic origins.

The prevailing approach to studying complex systems relies on mechanism-based model equations (11–20) (Supplementary Materials 1). While these models have proven effective within specific domains, they remain inherently constrained by the assumptions tied to their underlying mechanisms. Given the fundamental limitations of the reductive method (21, 22), there is widespread skepticism about the plausibility of formulating a general governing equation capable of capturing the behavior of complex systems across diverse domains (1, 23).

**The Complex Systems Response (CSR) Function**

Inspired by Newton's use of inductive reasoning employed before he formalized the laws of motion. We adopt a similar strategy to infer complex system behavior from empirical patterns. The inductive approach consists of four steps:

1. ***Observation***: Newton studied natural phenomena, including Galileo's experiments on falling objects and Kepler's observations of planetary motion.
2. ***Pattern Recognition***: In the *Principia*, Newton stated that "the change of motion (momentum) is proportional to the motive force impressed" (24,25).
3. ***Formulization***: Euler later expressed this pattern analytically as F=d(mV)/dt (26).
4. ***Generalization***: This law became a cornerstone of science and engineering, underpinning a wide range of applications.

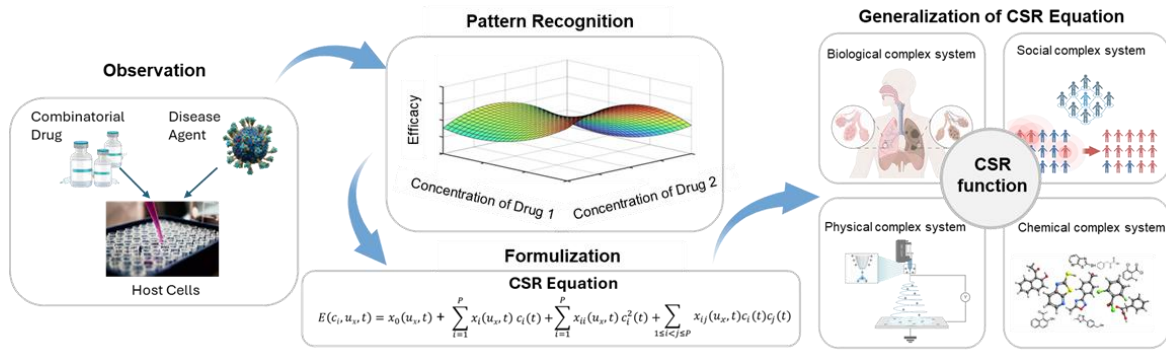

Figure 1: Inductive Approach Formulated Complex Systems Response (CSR) Function

1. ***Observation:*** In our previous study on a non-small cell lung cancer line treated with a combinatorial regimen, decreased cellular adenosine triphosphate (ATP) levels correlated with increased cancer cell death, marking an emergent system response (27).
2. ***Pattern Recognition***: Artificial neural network analysis revealed that ATP levels formed a smooth, multidimensional surface shaped across the entire ranges of drug-dose combinations.

3. **Formulation:** Regression analysis captured this surface as a second-order nonlinear Complex Systems Response (CSR) Equation (Equation-1), exhibiting the geometry of an elliptic or hyperbolic paraboloid (Supplementary Materials 2) (28-30). The CSR equation is rooted in observed natural patterns, inherently embedding the fundamental laws of nature within its mathematical formulation.

4. **Generalization:** This inductively derived Eunction-1 has since been generalized and validated across ~30 disease models, including cellular (31–35, 44), animal (36–37), and clinical studies (38–43, 49).

$$E(c_i, u_x, t) = x_0(u_x, t) + \sum_{i=1}^{P} x_i(u_x, t) \, c_i(t) + \sum_{i=1}^{P} x_{ii}(u_x, t) \, c_i^2(t) + \sum_{1 \leq i < j \leq P} x_{ij}(u_x, t) c_i(t) c_j(t) \qquad \text{.......(Equation-1)}$$

Here, $E(c_i, u_x, t)$ represents the emerging property, where $c_i(t)$ denotes the level of controllable parameter $i$, and $u_x$ represents unknown, uncontrollable or controllable parameters not included in the analysis. The coefficient $x_i(u_x,t)$ captures the system response to $c_i(t)$, while $x_{ii}(u_x,t)$ reflects the system response to $c_i^2(t)$, and $x_{ij}(u_x,t)$ characterizes the system responses to $c_i(t)$ and $c_j(t)$.

The CSR equation takes the simple form of a second-order nonlinear expression yet governs the complex behaviors of diverse systems through coefficients that dynamically adapt to the system's response to external stimulation. These adaptive couplings self-organize to produce the deterministic emergent property $E(c_i, u_x, t)$. Within this framework, the CSR equation redefines the third defining attribute of complex systems by establishing a direct quantitative relationship between the emergent property and the collective dynamics arising from interacting elements. The first attribute, interaction among components, is manifested through the coupling of parameters with the system's internal state. Furthermore, the CSR framework refines the second attribute by revealing the intrinsic second-order nonlinearity that fundamentally governs the behavior of complex systems.

In a complex system, the emergent property $E(c_i, u_x, t)$ depends on both the controllable parameters and their levels, $c_i(t)$. With P parameters, each at L levels, the search space expands to $L^P$ possible combinations, an intractable scale for exhaustive exploration. In contrast, the CSR equation contains only $(P^2 + 3P + 2)/2$ unknown coefficients. Calibration using a limited number of system-specific measurements enables deterministic evaluation of emergent properties across the entire *P*-dimensional parameter space (Supplementary Materials 3), eliminating the need for data-intensive statistical or AI-based methods (Supplementary Materials 4). Moreover, the inherent second-order nonlinearity of the CSR equation guarantees the existence of a single global optimum.

## Results

### *CSR Equation for Physical and Chemical Complex Systems*

The CSR equation, derived through mechanism-free inductive analysis, is inherently indication-agnostic. The main aim of this study is to demonstrate its broader applicability beyond biological systems, extending to physical, chemical and social complex systems.

To evaluate the generalizability of the CSR equation to physical and chemical systems, we chose Selective Laser Melting (SLM), a key technique in advanced manufacturing (51–53). In SLM, a laser scans a powder bed to form a transient melt pool that rapidly solidifies, enabling the fabrication of high-strength, high-resolution structures through precise control of the laser scanning path.

This study focuses on compressive yield strength, CYS, in stainless steel 316L as the emergent property. Among numerous SLM parameters, three parameters were selected for their direct influence on melting and solidification: laser power, $P$, scanning velocity, $V$, and powder bed temperature, $T$. The energy density, $E \sim P/V$ (Equation S4 in Supplementary Materials 5.1), quantifies the laser energy input per unit area and critically influences melt pool dynamics and CYS development. T governs the global thermal gradient and heat dissipation from the melt pool to the environment during solidification. Manufacturing and sample preparation details are provided in Supplementary Materials 5.

Here, $P$, $V$, and $T$ are treated as manufacturing parameters in Equation-2, while $E$ and $T$ serve as analysis parameters in Equation-3 to interpret experimental outcomes. This framework clarifies how these parameters collectively govern CYS.

$CYS( P, V, T, u_x) = x_0(u_x) + x_P(u_x) \times P + x_V(u_x) \times V + x_T(u_x) \times T + x_{PP}(u_x) \times P^2 + x_{VV}(u_x) \times V^2 + x_{TT}(u_x) \times T^2 + x_{PV}(u_x) \times P \times V + x_{VT}(u_x) \times V \times T + x_{TP}(u_x) \times T \times P$ .........(Equation-2)

$CYS( E, T, u_x) = x_0(u_x) + x_E(u_x) \times E + x_T(u_x) \times T + x_{EE}(u_x) \times E^2 + x_{TT}(u_x) \times T^2 + x_{ET}(u_x) \times E \times T$ ...............................................................................................................................................…..……(Equation-3)

For the three manufacturing parameter, P, V, and T, only ten specimens, $(3^2+3\times3+2)/2=10$, are required to determine the coefficients of Equation-2. After measuring the CYSs of 10 different parameter-level combinations, the ten coefficients in Equation-2 can be determined. With the measured CYSs, the six coefficients in Equation-3 are as follows:

$CYS = -7.54 \times 10 + 1.12 \times 10 \times E + 7.94 \times 10^{-1} \times T - 5.55 \times 10^{-2} \times E^2 - 3.59 \times 10^{-3} \times T^2$

$+ 7.38 \times 10^{-4} \times E \times T$...........................................................................................................(Equation-4)

The CSR surface, shown in Figure 1, exhibits concave elliptic paraboloid geometry. With an $R^2$ of 0.898, the CSR equation demonstrates strong predictive power of CYS. It validates the applicability of the CSR equation to the SLM process. The CSR surface visualizes CYS as a function of E and T capturing their influence across the entire multidimensional parameter space. The global maximum CYS of 545 MPa occurred at E=102 J/mm³ and T=121°C.

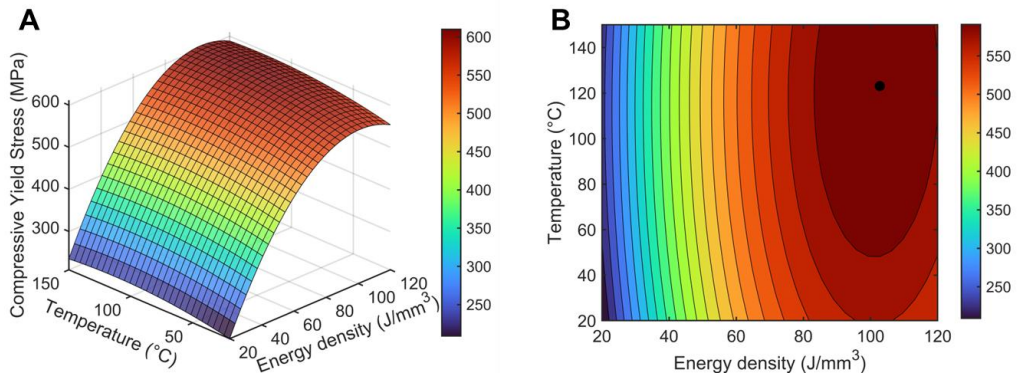

Figure 2. Specimens of stainless steel 316L were fabricated using SLM method. The energy density ranged from 20 to 120 J/mm³, and the powder bed temperature from 20 to 150 °C.

A. CYS is plotted as a function of E and T, yielding a $R^2$ value of 0.898.
B. A contour plot of CYS as a function of E and T. The black dot shows maximum value of CYS at 545 MPa at E=102 J/mm³ and T=121°C.

Notably, the maximum CYS obtained lies at the high end of the reported range, despite prior values being derived from a broad trial-and-error approach that required substantially more specimens and mechanical tests (54, 55, 56).

More studies further validate the applicability of the CSR equation to physical and chemical complex systems:

In a study evaluating tensile strength as a function of electrochemical plating frequency and Ni concentration, the CSR equation achieved $R^2 = 0.776$. The optimized NT-Cu-Ni foil exhibited a 25.4% increase in tensile strength over the baseline NT-Cu foil (57).

In three-dimensional electronic packaging, fully filling high-aspect-ratio through-glass vias (TGVs) with electrochemically plated copper is a major challenge. Using a CSR equation with five operating parameters ($R^2 = 0.90$), optimal parameter-level combinations successfully resolved this critical manufacturing issue (58).

Three comprehensive meta-analyses (Supplementary Materials 6) further validate the CSR equation across physical and chemical complex systems:

TiN thin-film coatings deposited by physical vapor deposition (PVD) exhibited strong predictive accuracy with the CSR equation, yielding $R^2 = 0.87$ for hardness, $0.99$ for scratch resistance, and $0.99$ for surface roughness (59). In a heat-transfer study using graphene nanofluids, the CSR equation ($R^2 = 0.99$) identified the minimum thermal resistance with only 13 calibration tests, matching the results of 214 exhaustive search trials reported in the original study (60). Likewise, in modeling composition–hardness relationships in high-entropy alloys (HEAs), the CSR equation ($R^2 = 0.93$) produced outcomes consistent with both experimental measurements and artificial neural-network (ANN) predictions (61). Collectively, these high $R^2$ values underscore the robustness and general validity of the CSR framework across physical and chemical complex systems.

### CSR Equation for Social Complex Systems

Based on data from multiple sources (62, 63), we investigated whether the numbers of social establishments, art galleries (AG), bookstores (BS), and restaurants (R), in different cities follow the CSR equation.

The self-organizations are affected by many influencing parameters, such as population size (P), city area (A), and gross domestic product (GDP) and others. In addition, the self-organization of each establishment, AG, BS, and R, was also influenced by distinct historical, cultural, educational, and economic factors, resulting in different system responses and intrinsic variability. These parameters are part of $u_x$ in Equation-1. Limited interactions among the establishments across cities further amplify these heterogeneities. As a result, statistical analyses often exhibit substantial data scatters, leading to the use of logarithmic transformations (64).

The CSR model using Log(P) alone yields a modest $R^2$ of 0.45, indicating that population alone does not adequately explain system behavior. Adding city area (Log(A)) increases $R^2$ to 0.79, and further including GDP per capita (Log(G/P)) raises it to 0.863 (Table 1), confirming that Log(P), Log(A), and Log(G/P) are dominating parameters of Log(AG). All three panels in Figure 3 show that cities with higher GDP per capita tend to host more art galleries. The substantial scatter, captured by the large Normalized Standard Deviation (NSD) of the logarithmic data, likely reflects cultural, historical, educational, and other city-specific contextual parameters.

| | CSR function | Function | $R^2$ | NSD |
|---|---|---|---|---|
| $C_1$: P | $\text{Log(AG)}= (-16.78) + 5.38\ \text{Log}(C_1) + (-0.38)\ \text{Log}(C_1)^2$ | 5 | 0.45 | 0.25 |
| $C_1$:P $C_2$:A | $\text{Log(AG)}= (-8.97) + (2.11)\ \text{Log}(C_1) + (0.08)\ \text{Log}(C_1)^2 + (1.27)\ \text{Log}(C_2) + (0.49)\ \text{Log}(C_2)^2 + (-0.72)\ \text{Log}(C_1)\text{Log}(C_2)$ | 6 | 0.79 | 0.1 |
| $C_1$:P $C_2$:A $C_3$:G/P | $\text{Log(AG)} = (-2.5) + (309)\ \text{Log}(C_1) + (-27)\ \text{Log}(C_1)^2 + (-109)\ \text{Log}(C_2) + (0.44)\ \text{Log}(C_2)^2 + (-323)\ \text{Log}(C_3) + (28.49)\ \text{Log}(C_3)^2 + (9.7)\ \text{Log}(C_2)\ \text{Log}(C_3) + (1.22)\ \text{Log}(C_1)\ \text{Log}(C_3) + (9.07)\ \text{Log}(C_1)\ \text{Log}(C_2)$ | 7 | 0.863 | 0.08 |

Table 1: CSR equations describing the number of art galleries of different cities expressed using one-, two-, and three-parameters

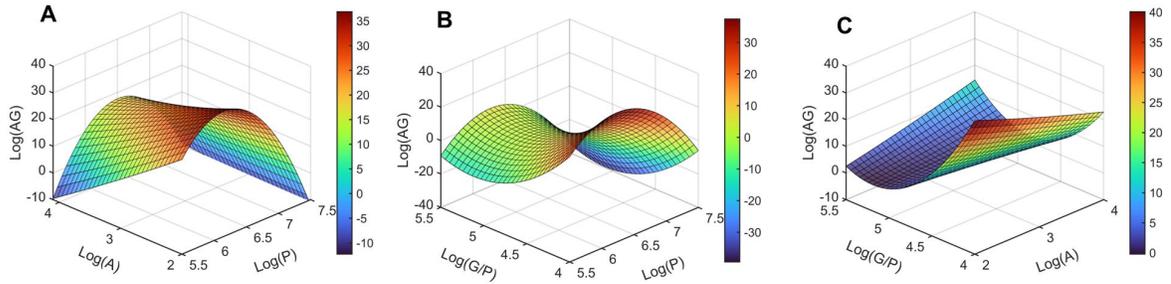

Figure 3: CSR surfaces of art gallery (AR) as a function of three parameters:

A. Art gallery as a function of city area (A) and population size (P)；

B. Art gallery as a function of gross domestic product per capita (GDP/P) and population size (P)；

C. Art gallery as a function of gross domestic product per capita (GDP/P) and City area (A).

Table 2 shows that Log(AG), Log(BS), and Log(R) scale with both population density, Log(P/A), and GDP per capita, Log(GDP/P). The resulting CSR surfaces fit their respective CSR equation well (Figure 4). Notably, these establishments are most concentrated in cities with either low or high, rather than intermediate, population densities, revealing a non-monotonic trend.

The single-parameter CSR equation can be expressed through logarithmic power-law relationships, which have been extensively investigated. Studies of power laws, whether based on linear dependence (65) or second-order dependence (66), have been widely reported across a variety of social complex systems. These findings reveal robust patterns, indicating that diverse social systems are governed by common organizing principles.

| Parameter | CSR function | Equation | $R^2$ | NSD |
|---|---|---|---|---|
| $C_1$:P/A $C_2$:G/P | $\text{Log(AG)}= (9.64) + (-4.15)\ \text{Log}(C_1) + (0.49)\ \text{Log}(C_1)^2 + (-0.02)\ \text{Log}(C_2) + (-0.13)\ \text{Log}(C_2)^2 + (0.25)\ \text{Log}(C_1)\ \text{Log}(C_2)$ | 8 | 0.8 | 0.1 |
| $C_1$:P/A $C_2$:G/P | $\text{Log(BS)}= (22.04) + (-6.23)\ \text{Log}(C_1) + (0.49)\ \text{Log}(C_1)^2 + (-3.05)\ \text{Log}(C_2) + (0.01)\ \text{Log}(C_2)^2 + (0.62)\ \text{Log}(C_1)\ \text{Log}(C_2)$ | 9 | 0.69 | 0.13 |
| $C_1$:P/A | $\text{Log(R)} = (41.2) + (-6.0)\ \text{Log}(C_1) + (0.59)\ \text{Log}(C_1)^2 + (-10.3)\ \text{Log}(C_2) + (0.76)\ \text{Log}(C_2)^2 + (0.49)\ \text{Log}(C_1)\text{Log}(C_2)$ | 10 | 0.77 | 0.1 |

| C$_2$ $^,$ GP | | | | | |
|---|---|---|---|---|---|

**Table 2:** The CSR equations of numbers of art galleries, bookstores and restaurants as function of GDP per capita (GDP/P) and population density (P/A)

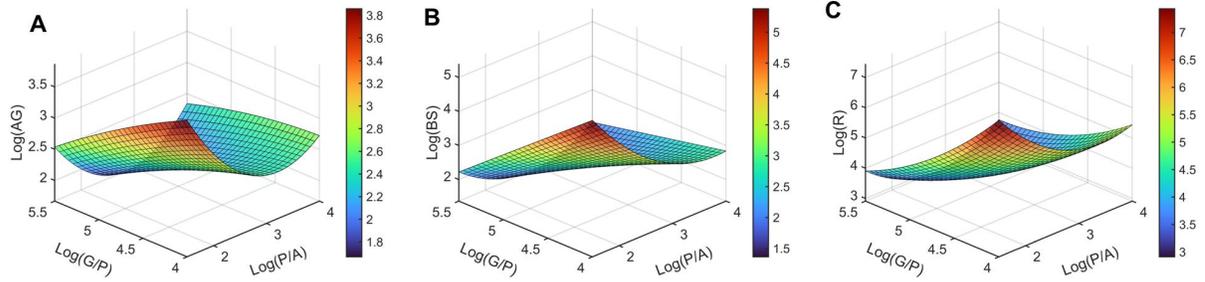

Figure 4: CSR surfaces of number of art galleries (AG), book stores (BS) and restaurants (R) as a function of two parameters: gross domestic product per capita (GDP/P) and population density (P/A)

A. Number of art galleries as a function of gross domestic product per capita (G/P) and population density (P/A)
B. Number of book stores as a function of gross domestic product per capita (G/P) and population density (P/A)
C. Number of art galleries as a function of gross domestic product per capita (G/P) and Population density (P/A)

### *Self-Organization Captured by the CSR Equation*

The CSR equation, derived through a mechanism-free, inductive approach, captures how interactions between stimulation parameters and system responses give rise to emergent properties across biological, physical, and social complex systems, spanning both homogeneous and heterogeneous regimes. Although deceptively simple, this domain-independent second-order nonlinear form governs a wide range of behaviors through coefficients that adapt dynamically as self-organization unfolds. The subsequent discussion elucidates the interplay between self-organizing dynamics and system responses that collectively shape emergent behavior.

#### *Self-Organization in Homogeneous Complex Systems*

In selective laser melting (SLM), the laser heats homogeneous SS316L stainless steel powder, inducing a phase transition at its melting point of 1400 °C. The process features extremely rapid cooling rates (~$10^6$ °C/s) (65), which restrict atomic diffusion distance and markedly increase nucleation site density. This rapid solidification initiates the first stage of self-organization, as metal atoms transition into hierarchical crystalline structures, grains, dislocations, and precipitates. Grain boundaries hinder dislocation motion, thereby enhancing CYS. This microscale self-organization is primarily driven by processing parameters, such as P, V, and the intrinsic properties of the metal powder (67-71).

During solidification, T governs the macroscale thermal gradient and influences heat dissipation in the second stage of self-organization. At low energy density, excessive heat loss leads to incomplete melting, porosity, and low CYS, referred to as underheating.

Conversely, high energy density impairs heat dissipation, resulting in overheating and bulk deformation.

Laser power and scan velocity regulate microscale melting and cooling, enhancing nucleation and CYS, while T controls macroscale thermal behavior. Despite differing mechanisms and length scales, the mechanism-free CSR equation integrates both self-organization stages. Optimal energy density for maximizing CYS occurs at the equilibrium point between underheating and overheating.

*Hierarchical Stages of Self-Organization in a Single Complex System*

In an in vitro study on Herpes simplex virus type 1 (HSV-1) suppression, combinatorial drugs entering cells selectively activate and self-organize a limited number of key pathways, such as STAT1 and S6, within the vast cellular network to block infection (72). Drug–cell interactions intensify over time, eventually reaching a critical phase, such that signals from individual cells interact with neighboring cells in tissues, generating new types of stimulation. These tissue-level responses, represented by the CSR equation's coefficients, differ from cellular-level responses and undergo higher-order self-organization, reaching a new critical phase before progressing to organ-level dynamics. As Anderson (1972) (22) noted, "At each level of complexity, entirely new properties appear, and understanding these behaviors requires research as fundamental as any other." Ultimately, these self-organizations and critical phases culminate in the whole-body response, $E(c_i, u_x, t)$ in Equation -1.

In these hierarchical stages of self-organization, the CSR equation applies at the homogeneous cell level and ultimately at the level of a single complex body system, such as a patient undergoing personalized therapy (38-41,43). It is important to note that while self-organization and critical transitions are distinct at each level, the CSR equation retains its same structural form, characterized by second-order nonlinearity, with different system response coefficients. This consistency provides a deterministic and quantitative foundation for therapeutic guidance across biological scales in the same complex system of the patient's body. It is interesting to note that an even higher level of complex system, the group of patients subjected to personalized therapy, also follows the CSR equation (Supplementary Materials 4, Figure S1).

*Self-Organization in Heterogeneous Complex Systems*

In a city, the number of social establishments, such as restaurants, is primarily influenced by parameters such as P/A and G/A. As these parameter changes, the number of establishments in the city undergoes reorganization until reaching a critical state of supply–demand balance. The long-term self-organization is shaped by a different set of city-specific parameters, including culture, history, and average education level. At a higher level of complexity, these heterogeneous systems of social establishments across different cities exhibit substantial scatter due to the long-term local parameters. The non-interacting complex systems are heterogeneous. Nonetheless, the scaling trends remain consistent with the CSR equation (Supplementary Materials 4-3).

While mechanism-based equations can model self-organization within a specific stage, they are limited in scope and fail to capture the diverse parameter, system interactions that occur across subsequent stages (11, 14, 15). In contrast, as discussed above, the mechanism-free CSR equation can track the process from the initial stimulus, bridge interactions across multiple stages of self-organizations, and ultimately link them to the system's emergent properties.

**Acknowledgments**

We thank Mr. Michael Jiunn Jye Chung for his invaluable assistance in facilitating experiments at the Centre for Additive Manufacturing, National University of Singapore. We are grateful to Professors Jien-Wei Yeh and An-Chou Yeh (National Tsing Hua University) and Professor Xiaoyu Cui (University of Shanghai for Science and Technology) for providing raw data that enabled our meta-analyses. We also thank Mr. Mingyuan Liu (Duke University) for preparing the figures.

**Funding**

National Additive Manufacturing Innovation Cluster grant 2017015) (JYHF)

UCLA Robert Benson Estate Fund grant R47441-AZ (CMH)

**Author contributions:**

Conceptualization: CMH, JYHF, LY

Methodology: CMH, JYHF, LY, JHK

Investigation: LY, JHK, CMH, JYHF, AK, WJC

Visualization: LY, JHK, CMH, JYHF, AK, WJC

Funding acquisition: JYHF, CMH

Project administration: LY

Supervision: JYHF, CMH

Writing – original draft: CMH, LY

Writing – review & editing: LY, AK, CMH

**Diversity, equity, ethics, and inclusion**

The diversity within our team, spanning multiple nationalities, ethnicities, and genders, enriched the perspectives and approaches brought to this research. We maintained an equitable collaborative environment, ensuring all members contributed meaningfully. This inclusive approach strengthened the scientific process and outcomes.

**Competing interests**

*Related Patent*

LY, JHK, CMH, and JYHF are co-inventors of the pending patent "Optimizing Process Parameters in Additive Manufacturing" (WO 2021/225529 A1).

*Conflict of Interest*

Authors declare that they have no competing interests.

**Data and materials availability**

All the data used for this manuscript are available in the text and supplementary materials. Additional supplementary data are available upon request from Dr. Lina Yan at Linayan46@outlook.com.

# Directly Mapping Interacting Components to Complex Systems' Emergent Properties

Lina Yan[1]†, Jeffrey Huy Khong[2]†, Aleksandar Kostadinov[3], Wen-Jun Chen[4]. Jerry Ying Hsi Fuh[1]*,

Chih-Ming Ho[1,5]*

Corresponding author: chihming@g.ucla.edu

## 1. Mechanism-Based Model Equations for Complex Systems

Mechanism-based model equations in complex systems are rooted in reductionist principles, employing mathematical formulations derived from well-defined physical, chemical, biological, or social mechanisms. Consequently, their validity is confined to the specific domains governed by those mechanisms and cannot be extended to different self-organization stages within the same system or to other complex systems governed by distinct mechanisms. Classic examples include the Ising, Fisher, and Black–Scholes models, each grounded in explicit mechanistic assumptions and validated within their respective domains through well-established equations that link components to their characteristic emergent properties.

By contrast, the CSR equation transcends mechanistic boundaries, providing a unified, mechanism-independent formulation that captures emergent behaviors across physical, chemical, biological, and social systems.

### 1.1: Ising Model

The Ising model (11), as expressed in Equation-S1, is a phenomenological model that describes discrete magnetic dipole moments (spins), taking values of +1 or −1, in a lattice. The spin configuration, denoted by σ. H(σ) is the system energy. The first-order terms correspond to the interaction of each spin with the external magnetic field, while the second-order terms represent pairwise interactions between spins at different lattice sites.

$$H(\sigma) = -\sum_i h_j\,\sigma_j - \sum_{i,j} J_{ij}\sigma_i\,\sigma_j \ \dots\dots\dots\dots\dots\dots\dots\dots\dots\dots\dots..\dots\dots\dots \text{(Equation-S1)}$$

The Ising model is a second order non-linear polynomial model equation for complex systems with bistable states. It has been widely applied across various domains, including magnetism (12) and spin glasses (13).

### 1.2: Fisher Model

The Fisher model (14) builds upon the statistical physics foundations to describe critical phenomena and phase transitions in physical complex systems. It links microscopic interactions to macroscopic observables through the partition function, which encapsulates the statistical properties of a system at thermal equilibrium. As the system approaches a critical point, singularities emerge in macroscopic quantities such as heat capacity, magnetization, and susceptibility, marking the onset of a phase transition. By applying the Renormalization Group (RG) framework, researchers elucidated how physical behavior transforms under successive scale transformations, revealing the universality of critical exponents across diverse systems. The Fisher formulation thus serves as a base in the modern understanding of critical phenomena, providing a quantitative bridge between microscopic interactions and emergent macroscopic order.

### 1.3 Black-Sholes Model

The Black–Scholes model, Equation-S2, is a widely used to help calculating a fair price for a specific type of financial contract called a European-style option, which gives someone the right to buy or sell stock at a fixed price in the future (15).

The key idea behind the model is that the value of an option can be closely estimated by combining two things:

1. the actual stock (or item being traded), and

2. a safe investment like a government bond.

By constantly adjusting the mix between these two, the value of the option can be tracked without taking on extra risk.

The Black–Scholes Partial Differential Equation (PDE) is as follows:

$$\partial V/ \quad \partial t + 1/2 \quad \sigma^2 S^2 \ \partial^2 V/\partial S^2 \quad + rS \ \partial V/\partial S \quad - rV = 0 \quad \ldots\ldots\ldots\ldots\ldots(\text{Equation–S2})$$

Where:
- $V(S,t)$ is the option price,
- $S$ is the underlying asset price
- $t$ is the time to maturity or the current time
- $r$ is the risk-free interest rate,
- $\sigma$ is the volatility of the underlying asset.

### 1.4 Response Surface Method (RSM)

The Response Surface Method (RSM) (16) is different from the above-mentioned physical-mechanism based equation. It is constructed from expanding a function $\eta$ into a polynomial series of parameters, $\chi_i$ .

$$\eta = \beta_o + \beta_1\chi_1 + \beta_2\chi_2 + \ldots + \beta_{11}\chi_1^2 + \beta_{22}\chi_2^2 + \ldots + \beta_{12}\chi_1\chi_2 + \ldots\beta_{111}\chi_1^3 + \ldots \quad \ldots\ldots\ldots\ldots(\text{Equation-S3})$$

In all complex systems, the preservation of causal relationships (17) is a fundamental requirement. Mathematics, by itself, has no inherent notion of causality. Nevertheless, RSM disregards this cause–effect principle and applied across scientific and engineering domains, often resulting in critical limitations. As shown by Myers *et al.* (18), optimization performed on non-causal response surfaces can converge to parameter combinations lacking physical meaning, resulting in unstable or misleading predictions in practice.

When the response function $\eta$ is expanded into a polynomial series, the coefficients $\beta_i$, $\beta_{ii}$, and $\beta_{ij}$ must remain constant to preserve mathematical consistency (Equation-S3). Otherwise, each coefficient would need to be further expanded as a function of its underlying parameters. To circumvent this inherent contradiction, the Response Surface Method (RSM) abandons the fundamental mathematical requirement of coefficient constancy and replaces it with statistical regression, thereby reducing these coefficients to numerical artifacts detached from the system's governing principles. Consequently, such regression-based surfaces merely approximate experimental data while violating both the physical law of causality and the mathematical constraint of coefficient constancy (19).

Moreover, as emphasized by Draper and Lin (20), second-order polynomial approximations with fixed coefficients are mathematically justifiable only within a narrow, local region of the factor space; beyond that domain, the model rapidly loses validity and may yield erroneous or physically inconsistent predictions. Collectively, these issues demonstrate that RSM, while statistically convenient, compromises fundamental physical and mathematical principles, often leading to unreliable or non-causal optimization outcomes.

## 2. Stages of Developing Complex Systems Response (CSR) Equation

### 2.1: Precursor of CSR Equation - Feedback System Control (FSC) Platform

Combinatorial treatments are routinely employed in managing a wide range of diseases; however, determining the optimal doses remains a major challenge. For a regimen involving $P$ drugs each at $L$ possible dose levels, the search space expands exponentially to $L^P$ possible combinations, rendering exhaustive searches impractical.

Prior to the discovery of the Complex Systems Response (CSR) function, we developed a Feedback System Control (FSC) approach that integrates feedback principles with game theory. FSC significantly reduces the search burden, identifying optimal drug-dose regimens in fewer than 20 feedback iterations, far more efficient than exhaustive $L^P$ searches. This methodology has been experimentally validated through many in vitro studies across diverse disease models (28-30).

### 2.2: The CSR Equation

In the investigation of diseased biological systems under multidrug treatment, we sought to study how therapeutic efficacy varies with dosage combinations. Operating in a high-dimensional parameter space without prior knowledge of the response surface topology, artificial neural networks (ANNs) provided a practical modeling approach. With only limited amount of training data, the ANNs consistently revealed a smooth and continuous response surface (27).

Successive validations, from extensive cell-line studies to animal models and ultimately clinical trials, yielded consistently high coefficients of determination, $R^2$, values near 1, confirming that the system behavior can be accurately represented by the Complex Systems Response (CSR) equation, equation-1, which takes the form of a second-order nonlinear polynomial (31- 43).

The coefficients of the CSR equation quantify how a specific system responds to combinations of input parameters and are therefore inherently system specific. While complex systems may involve numerous controllable parameters, in practice, a subset of these typically governs the self-organization process that leads to the desired therapeutic outcome (70). Moreover, unknown variables or controllable parameters excluded from the analysis, denoted as $u_x$, also contribute to system behavior. As a result, the CSR coefficients are not constant; they dynamically adapt with respect to $u_x$ and time. These adaptive coefficients allow the CSR equation to deterministically and quantitatively capture emergent system properties arising from parameter–system interactions. Through examinations of the properties of coefficients as studied in references (*31*) and (*37*), the generalized form of the CSR equation is presented as Equation-1.

Many names have been associated with the CSR platform. In the early days, flexible or streamlined Feedback System Control (s-FSC) were utilized (28, *31-34, 36*). Later, the Phenotypic Response Surface (PRS) equation replaced FSC or s-FSC for applications in biological complex systems (*38, 43*). Other research groups used CURATE.AI (*39, 40*), QPOP (35), and IDentif.AI (44). Eventually, the CSR equation was adopted after the discovery that Equation-1 could serve as a general transfer function for physical, chemical, biological and social complex systems.

## 3. Method of Implementing CSR Function

The Complex Systems Response (CSR) equation links the emergent property (outcome) to interacting components, input parameters ($P$) and their levels ($L$), through a response surface shaped as either an elliptic or hyperbolic paraboloid. The unknowns in the CSR equation are its coefficients, which are functions of $u_x$ and time. These coefficients are determined through calibration experiments.

$$E(c_i, u_x, t) = x_0(u_x, t) + \sum_{i=1}^{P} x_i(u_x, t)\, c_i(t) + \sum_{i=1}^{P} x_{ii}(u_x, t)\, c_i^2(t) + \sum_{1 \le i < j \le P} x_{ij}(u_x, t) c_i(t) c_j(t) \qquad \text{.........(Equation -1)}$$

### 3.1: Protocol for CSR-Based Steady System Optimization:

The CSR equation has a second order nonlinearity and one global optimal outcome.

*(1) Determine the Number of Calibration Tests:*

Using the CSR equation (Function-1), calculate the required number of calibration tests as ($P^2+3P+2$)/2, where $P$ is the number of input parameters.

*(1) Define Parameter Ranges:*

Specify the minimum and maximum levels for each parameter.

*(1) Design Experimental Matrix:*

To strategically distribute a limited number of parameter-level combinations across a multidimensional space, employ Orthogonal Array Composite Design (OACD) (45, 46). OACD ensures uniform coverage of the $P$-dimensional space with ($P^2+3P+2$)/2 test points, enabling accurate mapping of the CSR surface.

*(2) Conduct Calibration Experiments:*

Experimentally measured system outcomes correspond to the OACD-designed ($P^2+3P+2$)/2 combinations.

*(3) Determine CSR Coefficients:*

Use linear regression to solve for the ($P^2+3P+2$)/2 coefficients in the CSR function.

*(4) Identify Optimal Outcome:*

Analyze the CSR equation to determine the global optimal outcome and the corresponding parameter-level combination.

### 3.2: Protocol for Dynamic CSR-Based Unsteady System Optimization:

*(1) Determine the Number of Calibration Tests:*

Using the CSR equation (Equation -1), calculate the required number of calibration tests as M=($P^2+3P+2$)/2, where $P$ is the number of input parameters.

*(2) Define Parameter Ranges:*

Specify the minimum and maximum levels for each parameter.

*(3) Design Experimental Matrix:*

Employ Orthogonal Array Composite Design (OACD) (45, 46) to strategically distribute M parameter-level combinations across the multidimensional space.

*(4) Sequential Testing:*

Apply the OACD-designed parameter-level pairs sequentially over time.

*(5) Determine Optimal Dose:*

After completing the M tests, use the outcomes of $1^{st}$ test to $M^{th}$ test results to predict the optimal parameter-level combinations for the $(M+1)^{th}$ test.

*(6) Dynamic Adaption:*

Real-time optimization can be achieved through a moving-window approach, where the outcomes from a fixed set of recent tests, such as using results from the $2^{nd}$ through the $(M+1)^{th}$ tests are employed to determine the optimal parameter combination for the $(M+2)^{th}$ test and so on. This moving window continuously advances as new data becomes available, enabling the optimization process to adapt dynamically to system changes over time.

## 4. Optimization of Complex Systems

### 4.1 Optimization of Homogeneous Complex System

Elements used in material processing or cells used in vitro tests are all similar and are homogenous complex systems. The optimization process can be operated in parallel at the same time with $(P^2+3P+2)/2$ test samples of different parameter-level combinations (Supplementary Materials section 3.1). This efficient global optimization method has been validated in over 40 papers in physical and biological complex systems.

The response surface of Equation-1 assumes an elliptic paraboloid form when all eigenvalues are positive or all negative, and a hyperbolic paraboloid form when the eigenvalues have mixed signs.

### 4.2 Optimization of an Individual Complex System

The Complex System Response (CSR) equation enables patient-specific calibration with only $(P^2+3P+2)/2$ tests, sufficient to determine the system-response coefficients of an individual's dynamic physiology. Because each patient's physiological response is intrinsically unique and continuously varying, dosing must be dynamically adjusted in series over time according to the CSR-based optimization protocol described in Supplementary Materials Section 3.2. The CSR equation, characterized by second-order nonlinearity, ensures the identification of a global optimum in dosing. This capability frequently reveals therapeutic features that remain inaccessible under conventional approaches (43).

For example, in a clinical study of 10 HIV patients (41) the CSR method identified optimal maintenance doses that differed by nearly twofold among individuals, underscoring the critical importance of individualized system optimization in personalized therapy. In contrast, conventional cancer treatments rely on population-average dosing, which often produces suboptimal outcomes due to patient heterogeneity (47, 48). In one metastatic prostate cancer case, the personalized dose was approximately half of the FDA-recommended level (39). Likewise, in a 60-patient liver transplant trial, the CSR-guided group experienced accelerated recovery of liver function and achieved hospital discharge one-third earlier than the control cohort (43). Beyond pharmacological interventions, CSR-guided optimization has also been extended to cognitive training, where training parameters and levels were tailored to the dynamics of individual learning (40).

### 4.3 Optimization of Heterogeneous Complex Systems

The CSR-based personalized therapy (43) enables each patient, a single complex system, to achieve faster normalization of liver AST levels, resulting in shorter lengths of hospital stay (LOS). At an even higher level of complexity, the AST–LOS correlation across all personalized-treated patients, representing a collection of non-interacting, heterogeneous complex systems, continues to conform to the CSR equation (Equation-S3). However, because there are no interactions among patients, the logarithmic normalized standard deviation (NSD) becomes markedly high, with moderate $R^2$ values and large data scatter, indicating the emergence of numerous local maxima in the response surface. This behavior reflects the intrinsic loss of collective coherence when individual systems evolve independently, resulting in fragmented optima rather than a unified global response. (Figure-S1).

Log(LOS) = 0.67 + 0.47 Log(DAST)  …………………………………………… Equation-S3

$R^2$= 0.36 and NSD= 0.19

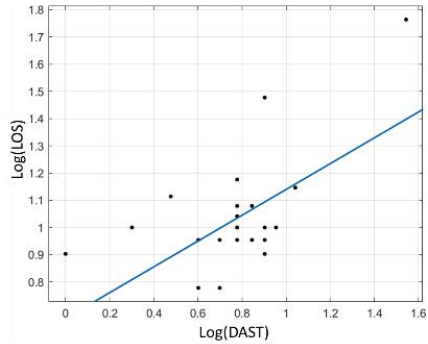

Figure S1: For the group of liver transplant patients treated with personalized therapy, the length of staying in the hospital (LOS) is plotted against the day from transplant data to the day when AST reached normal value (DAST).

In social complex systems, establishments such as bookstores, art galleries, and restaurants exhibit scaling behaviors consistent with the CSR equations (Table 2). However, these entities across different cities function as heterogeneous, non-interacting complex systems. Consequently, their logarithmic normalized standard deviations (NSDs) remain high (Table 2). Such large deviations arise because each establishment type evolves under distinct historical, socioeconomic, and environmental conditions, including traffic patterns and other uncontrollable parameters, collectively represented by $u_x$ in Equation 1. As a result, the response surfaces display numerous local optima, reflecting the intrinsic difficulty of achieving further optimization in the absence of intersystem interactions.

### 4.4 CSR Equation vs Big Data Based Optimization Methods

Many widely used optimization methods, such as Simulated Annealing, Genetic Algorithms, Particle Swarm Optimization, Monte Carlo Optimization, Bayesian Analysis, or AI-based search, require very big datasets, often numbering in the thousands, millions, or more. By contrast, the CSR method requires only $(P^2+3P+2)/2$ data points to determine the system-response coefficients and identify the global optimal emergent property.

The collection of such large datasets is both costly and time-consuming. By contrast, the outcome of the CSR equation resides on a well-defined CSR surface in multi-dimensional space (Figure-S2). Whereas the datasets collected from different sources and employed by statistical optimization methods are most likely heterogeneous. Consequently, most of the data points scatter across the search space. The ones not on the CSR surface are simply noises giving rise to many apparent local optima (Figure S1 and S2). Hence, simply enlarging the dataset does not guarantee convergence; rather, it often increases the standard deviation of the results.

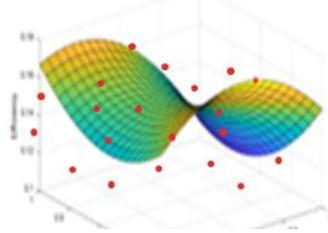

Figure S2: A sketch illustrating the outcomes of the CSR equation (the response surface) compared with those (red dots) from large-dataset statistical or AI-based methods."

Artificial Intelligence (AI) and Generative AI (GAI) are rooted in advances in machine learning and language-based technologies. Modern AI systems are predominantly powered by the transformer architecture, which forms the backbone of most GAI platforms. Transformers enable the establishment of input–output mappings by learning statistical relationships within vast datasets, thereby generating outputs aligned with the desired tasks.

Recently, the interdisciplinary applications of AI and GAI have been driving rapid scientific and technological breakthroughs across diverse domains, including engineering, materials discovery, medical diagnosis, and drug development. Current transformer-based models are primarily trained through unstructured brute-force extraction of statistical results from massive, heterogeneous datasets. This training paradigm demands computationally intensive workflows, involving high-dimensional optimization across billions of parameters (50).

The CSR and AI frameworks represent complementary approaches to understanding and optimizing complex systems. Not all tasks require big-data–driven AI. In contrast, most engineering and medical applications, except for highly data-intensive domains such as image processing and protein structure prediction, can be effectively characterized by only a few key governing parameters, typically one to five. For such systems, the CSR equation provides a general, deterministic input–output framework that captures system behavior with minimal calibration. Relying on big-data AI methods in these contexts is neither cost-effective nor computationally efficient, whereas CSR offers a small-data, physics-consistent alternative capable of achieving accurate and interpretable solutions.

## 5. Material and Manufacturing of Selective Laser Melting (SLM)

### 5.1 Sample Preparation

Stainless steel 316L specimens were fabricated using Selective Laser Melting (SLM). Each specimen, conforming to ASTM E9, had a cylindrical geometry with a diameter and height of 5 mm. Fabrication was conducted on an EoS M290 system (EoS, Germany) using recycled 316L stainless steel powder (Retsch GmbH, Germany) with a particle size of 50–80 μm.

Key manufacturing parameters were systematically varied: laser power (P) from 80–190 W, scanning velocity (V) from 100–2500 mm/s, and powder bed temperature (T) from 25–160 °C. Oxygen levels were maintained below 0.2% to minimize oxidation. A bi-directional scanning pattern with 67° rotation per layer ensured uniform material distribution.

Laser energy density (E), representing energy input per unit volume, was calculated as:

E = P / (V × h × l)   …………………………………………………………………...Equation-S4

where P, V, h, and l denote laser power, scan speed, hatch distance, and layer thickness, respectively. Default settings included a hatch distance (h) of 0.09 mm, layer thickness (l) of 0.02 mm, and a laser spot size of 0.08 mm.

After printing, specimens were detached from the build platform via wire cutting. To relieve residual stress, they were furnace-heated from ambient temperature to 400 °C over one hour, soaked for 4 hours, and furnace-cooled to room temperature.

### 5.2 Mechanical tests

The compressive tests were conducted using Instron 8848 (Instron, USA) following the ASTM E9 test protocol. An adjustable bearing block made of tungsten carbide was employed to evenly distribute the initial load. The compressive yield stress was recorded at the strain offset at 0.2%.

### 5.3 Area Density Function Measurement

The internal structure was revealed on planes perpendicular (XY) and parallel (XZ or YZ) to the build direction. Surfaces were ground and polished to a mirror finish, then electrolytically etched with oxalic acid following ASTM E3-11 and E407 standards. Optical images were captured on both planes using a Nikon Eclipse LV150N (Nikon, Japan) in bright field mode at 100× magnification over a 1 mm × 1 mm area (Figure-S3).

Porosity within the samples was quantitatively evaluated using digital image analysis. Images from both XY and XZ planes were processed to differentiate pores from the solid matrix based on pixel contrast. The porosity percentage was calculated using ImageJ, an open-source image analysis software developed by the U.S. National Institutes of Health (https://imagej.net/ij/index.html).

The volumetric area fraction of density (VAFD) was estimated using a geometric approximation that combines porosity measurements from the XY and XZ planes. This method assumes that the volume fraction of solid material can be inferred from the product of the area fractions in two orthogonal sections. The VAFD was calculated as:

VAFD=(1−$P_{XY}$)×(1−$P_{XZ}$) …………………………………………………………Equation-S5

where $P_{XY}$ and $P_{XZ}$ are the porosity area fractions measured on the XY and XZ planes, respectively.

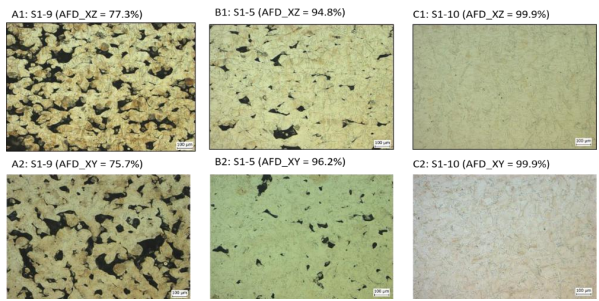

**Figure S3. Demonstration of VAFD for Porous Specimens S1-9 and S1-5, and Solid Specimen S1-10**

The volumetric area fraction of density (VAFD) was calculated using the relation:

VAFD$=(1-P_{XY})\times(1-P_{XZ})$ where $P_{XY}$ and $P_{XZ}$ represent porosity fractions on the XY and XZ planes, respectively (see Table S1). Microscopic images for each sample are shown on both XY and XZ planes.

(A) For sample S1-9, the porosities on the XY and XZ planes were 75.7% and 77.3%, respectively, resulting in a VAFD of 58.5%.

(B) For sample S1-5, porosities were 96.2% (XY) and 94.8% (XZ), yielding a VAFD of 91.2%.

(C) For the solid sample S1-10, both planes showed 99.9% porosity, corresponding to a VAFD of 99.8%.

### 5.4 List of experimental data

For an experiment involving three parameters pressure (P), volume (V), and temperature (T), the CSR equation requires ten calibration tests. However, certain parameter combinations can result in defective specimens due to unforeseen manufacturing issues. To mitigate this, 60% more specimens were fabricated than the minimum required, ensuring a sufficient number for mechanical testing. Of the 16 specimens initially printed (S1-1 to S1-16), five were excluded from the final analysis. Two exhibited energy densities are too low to melt the powder, producing very porous specimens unsuitable for mechanical testing. The other three had excessively high energy densities, causing over-melting and severe distortion. The remaining 11 specimens, along with their process parameters and compressive yield stress values, are presented in Table S1.

| Specimen | Power P (W) | Velocity V(mm/s) | Temperature T (°C) | Energy density E (J/mm³) | Compressive Yield Stress CYS (MPa) | Volumetric Area Fraction of Density VAFD (%) |
|---|---|---|---|---|---|---|
| S1-1 | 190 | 2000 | 160 | 52.8 | 437.1 | 96.4 |
| S1-4 | 80 | 400 | 160 | 111.1 | 534.2 | 99.8 |
| S1-5 | 190 | 2000 | 25 | 52.8 | 413.1 | 91.2 |
| S1-8 | 80 | 400 | 25 | 111.1 | 534.1 | 99.6 |
| S1-9 | 80 | 1400 | 70 | 31.7 | 243.1 | 58.5 |
| S1-10 | 120 | 800 | 70 | 83.3 | 554.3 | 99.8 |
| S1-12 | 120 | 1400 | 160 | 47.6 | 302.6 | 89.6 |
| S1-13 | 120 | 1400 | 25 | 47.6 | 361.3 | 84.0 |
| S1-14 | 160 | 1400 | 115 | 63.5 | 474 | 90.2 |
| S1-15 | 160 | 900 | 25 | 98.8 | 446.2 | 99.8 |
| S1-16 | 120 | 2000 | 70 | 33.3 | 249.6 | 54.7 |

Table S-1: Summary of process parameters and compressive yield stress, and Volumetric Area Fraction of Density

## 6 Meta-analysis of physical and chemical complex Systems

### 6.1 PVD TiN thin film coating for higher coating hardness and adhesion (59)

**Abstract:** An optimization study was conducted on the parameters of titanium nitride (TiN) coating deposited on aerospace-grade Al7075-T6 alloy using the magnetron sputtering technique. The effects of substrate temperature, DC bias voltage, nitrogen flow rate, and DC power on surface hardness, adhesion, surface roughness, and microstructure of the coated samples were systematically investigated.

**Parameters and Levels in Reference (57):**

| Parameters | DC power (W) | Temperature (°C) | Nitrogen flow rate (%) | Substrate bias voltage (V) |
|---|---|---|---|---|
| Range of levels | 300, 350, 400, 500 | 150, 180, 200, 220 | 3, 4, 5, 6 | 25, 50, 75, 100 |

Table S-2: Parameters and Levels

**Test Results in Reference (57):**

16 tests were performed. (Data: Table 3 in the paper (57))

| Run | DC power (W) | Temperature (°C) | Nitrogen flow rate (%) | Substrate bias voltage (V) | Hardness (HV) | Scratch force (mN) | Roughness (μm) |
|---|---|---|---|---|---|---|---|
| 1 | 300 | 150 | 3 | 25 | 204 | 708 | 0.051 |
| 2 | 300 | 180 | 4 | 50 | 210 | 723 | 0.056 |
| 3 | 300 | 200 | 5 | 75 | 201 | 850 | 0.073 |
| 4 | 300 | 220 | 6 | 100 | 720 | 418 | 0.047 |
| 5 | 350 | 150 | 4 | 75 | 180 | 1404 | 0.041 |
| 6 | 350 | 180 | 3 | 100 | 185 | 713 | 0.059 |
| 7 | 350 | 200 | 6 | 25 | 285 | 1458 | 0.204 |
| 8 | 350 | 220 | 5 | 50 | 190 | 1030 | 0.035 |
| 9 | 400 | 150 | 5 | 100 | 410 | 2446 | 0.056 |
| 10 | 400 | 180 | 6 | 75 | 200 | 314 | 0.066 |
| 11 | 400 | 200 | 3 | 50 | 400 | 1611 | 0.123 |
| 12 | 400 | 220 | 4 | 25 | 720 | 1500 | 0.058 |
| 13 | 500 | 150 | 6 | 50 | 240 | 1071 | 0.117 |
| 14 | 500 | 180 | 5 | 25 | 300 | 2482 | 0.068 |
| 15 | 500 | 200 | 4 | 100 | 410 | 1977 | 0.064 |
| 16 | 500 | 220 | 3 | 75 | 235 | 715 | 0.289 |

Table S-3: Test Results in the paper (57)

**Optimal Outcome in Reference (57):**

| | DC Power (W) | Temperature (ºC) | Nitrogen Flow Rae (5) | Substrate Bias Voltage (V) | Optimal output |
|---|---|---|---|---|---|
| Hardness | 400 | 220 | 4 | 100 | 840 Hv |
| Scratch force | 500 | 200 | 5 | 75 | 2,588 mN |
| Roughness | 300 | 150 | 5 | 25 | 0.0317 μm |

Table S-4: Optimal Outcome analysis in the paper (57)

**Meta-Analysis**

For three parameters, 15 parameter-level tests will be able to determine the maximum outcome and its optimal parameter-level combinations

**Ranges of the parameter levels**

|  | DC power (W) | Temperature (°C) | Nitrogen flow rate (%) | Substrate bias voltage (V) |
|---|---|---|---|---|
| Range of levels | 300-500 | 150-220 | 3-6 | 25-100 |

Table S-5: Ranges of the parameter levels

**Data:** For four parameters need 15 tests. The 16 tests in the Table 3 of the refernce (57) were used.

**CSR Function:**

1. Hardness = (-307.39) + (-1.47) $c_1$ + (0.02) $c_1^2$ + (26.18) $c_2$ + (2e-03) $c_2^2$ + (-941.17) $c_3$ + (54.28) $c_3^2$ + (4.38) $c_4$ + (0.11) $c_4^2$ + (4.08) $c_3c_4$ + (-0.21) $c_2c_4$ + (1.76) $c_2c_3$ + (4e-03) $c_1c_4$ + (-0.47) $c_1c_3$ + (-0.05) $c_1c_2$

$R^2$=0.87

2. Scratch force = (-3367.14) + (-38.32) $c_1$ + (0.19) $c_1^2$ + (264.81) $c_2$ + (-0.49) $c_2^2$ + (-10312.07) $c_3$ + (346.85) $c_3^2$ + (343.63) $c_4$ + (-0.34) $c_4^2$ + (13.68) $c_3c_4$ + (-2.23) $c_2c_4$ + (44.42) $c_2c_3$ + (0.12) $c_1c_4$ + (-6.53) $c_1c_3$ + (-0.43) $c_1c_2$

$R^2$=0.99

3. Roughness = (0.03) + (-1e-03) $c_1$ + (6.92e-06) $c_1^2$ + (2e-03) $c_2$ + (5.17e-06) $c_2^2$ + (0.09) $c_3$ + (0.04) $c_3^2$ + (7e-03) $c_4$ + (-5.24e-06) $c_4^2$ + (-1e-03) $c_3c_4$ + (-3.37e-05) $c_2c_4$ + (-1e-03) $c_2c_3$ + (3.23e-06) $c_1c_4$ + (-1e-03) $c_1c_3$ + (4.79e-07) $c_1c_2$
$R^2$=0.99

**Outcome:  CSR equation based optimization:**

|  | DC Power (W) | Temperature (ºC) | Nitrogen Flow Rae (5) | Substrate Bias Voltage (V) | Optimal outcome |
|---|---|---|---|---|---|
| Hardness | 500 | 220 | 3 | 25 | 1050.2 Hv |
| Scratch force | 500 | 150 | 3 | 100 | 13,328 mN |
| Roughness | 456 | 20 | 5.7 | 100 | 0 μm |

Table S-6: CSR Equation Based Optimal Outcome

**Conclusion:**

The CSR platform is validated in this complex system with a high coefficient of determination ($R^2$). The CSR surface maps the distribution of material properties across the entire P-dimensional space and identifies the global optimum, which surpasses the suboptimal values previously reported in the literature (59)

|  | CSR based meta-analysis | | | | Results reported in the paper (57) | | | |
|---|---|---|---|---|---|---|---|---|
| Hardness | 1050.2 HV | | | | 840 HV | | | |
| Opt. parameters | 500W | 220°C | 3% | 25V | 400W | 220°C | 4% | 100V |
| Scratch force | 13,328 mN | | | | 2,588 mN | | | |
| Opt. parameters | 500W | 150°C | 3% | 100V | 500W | 200°C | 5% | 75V |
| Roughness | 0 μm | | | | 0.0317μm | | | |
| Opt. parameters | 456W | 220°C | 5.7% | 100V | 300W | 150°C | 5% | 25V |

Table S-7: The CSR based global optimal hardness, scratch force, roughness and the corresponding material properties in this meta-analysis paper.

## 6.2 Heat transfer of an oscillating heat pipe with graphene nanofluids (60)

**Reference:** Yu Zhou,  Xiaoyu Cui, Jianhua Weng, Saiyan Shi, Hua Han,  Chengmeng Chen, "Experimental investigation of the heat transfer performance of an oscillating heat pipe with graphene nanofluids", Powder Technology ,v. 332, 371–380, https://doi.org/10.1016/j.powtec.2018.02.048 , 2018

**Abstract:** The heat transfer performance of oscillating heat pipes (OHPs) using graphene nanoplatelet (GNP) nanofluids was investigated experimentally. Results show that OHPs charged with GNP nanofluids exhibit enhanced heat transfer performance compared to those using deionized water (DW)

**Parameters and Levels in Reference (58):**

| Parameters | GNP concentration (wt.%) | Filling ratios (%) | Heating power (W) |
|---|---|---|---|
| Range of levels | 0.02-0.167 | 45-90 | 10-100 |

Table S-8: Ranges of Parameter Levels

**Test Results:** 214 tests were performed. (Data: Contact information of the original author, Professor Xiaoyu Cui at University of Shanghai for Science and Technology, email：xiaoyu_cui@usst.edu.cn)

**Outcome in Reference (58):**

| | GNP concentration (wt.%) | Filling ratios (%) | Heating power (W) | Optimal outcome ºC/W |
|---|---|---|---|---|
| Thermal Resistance | 0.02-0.167 | 62 | 100 | 0.01 |

Table S-9: Optimal Outcome analysis in the paper (58)

**Meta-Analysis**

**CSR Platform:** For three parameters, 13 parameter-level tests will be able to determine the best outcome and its optimal parameter-level combinations.

**CSR Function:**

Thermal resistance = $8.81 - 30.1\,c_1 - 0.24\,c_2 - 0.01\,c_3 + 48.6\,c_1^2 + 0.02\,c_2^2 + 9.43\text{e-}5\,c_3^2 - 0.0004\,c_2 c_3 - 0.035\,c_1 c_3 + 0.39\,c_1 c_2$, $R^2 = 0.99$

**Outcome: CSR equation based optimization:**

| | GNP concentration (wt.%) | Filling ratios (%) | Heating power (W) | Optimal outcome ºC/W |
|---|---|---|---|---|
| **Thermal Resistance** | **0.63** | **71.2** | **100** | **0** |

Table S-10: CSR equation Based Optimal Outcome

**Conclusion:** With only 13 parameter-level calibration tests, the CSR equation ($R^2 = 0.99$) accurately predicts the minimum thermal resistance, consistent with the result obtained from 214 exhaustive search tests in the original study

## 6.3 Prediction of the Composition and Hardness of High-Entropy Alloys by Machine Learning (61)

**Reference:** Chang, Y. J., Jui, C. Y., Lee, W. J., & Yeh, A. C. (2019). Prediction of the composition and hardness of high-entropy alloys by machine learning. *Jom*, *71*(10), 3433-3442., https://doi.org/10.1007/s11837-019-03704-4, 2019

**Abstract:** This study employs an artificial neural network (ANN) to predict the composition of high-entropy alloys (HEAs) based on the non-equimolar AlCoCrFeMnNi system, with the aim of maximizing hardness. To further optimize the composition, the ANN is integrated with a simulated annealing algorithm.

**Parameters and Levels in Refence (59):**

| Parameters | Co (at%) | Cr (at%) | Fe (at%) | Ni (at%) | Mn (at%) | Al (at%) | Cu (at%) | Mo (at%) |
|---|---|---|---|---|---|---|---|---|
| Range of levels | 0 - 30.8 | 0 - 35 | 0 - 34.9 | 0 - 48.8 | 0 - 23.3 | 0 - 75 | 0 - 20 | 0 - 20 |

Table S-11: Ranges of Parameter Levels

**Test Results in Reference (57)**

97 tests were performed.  2 (Original data was provided by Professors An-Chou Yeh and Jien-Wei Yeh, National Tsinghua University)

| Run | Co (at%) | Cr (at%) | Fe (at%) | Ni (at%) | Mn (at%) | Al (at%) | Cu (at%) | Mo (at%) | Hardness (Hv) |
|---|---|---|---|---|---|---|---|---|---|
| 1 | 0.0000 | 0.0000 | 0.0000 | 25.0000 | 0.0000 | 75.0000 | 0.0000 | 0.0000 | 300 |
| 2 | 0.0000 | 24.3902 | 24.3902 | 48.7805 | 0.0000 | 0.0000 | 0.0000 | 2.4390 | 228 |
| 3 | 0.0000 | 23.8095 | 23.8095 | 47.6190 | 0.0000 | 0.0000 | 0.0000 | 4.7619 | 238 |
| 4 | 0.0000 | 23.2558 | 23.2558 | 46.5116 | 0.0000 | 0.0000 | 0.0000 | 6.9767 | 239 |
| 5 | 0.0000 | 22.7273 | 22.7273 | 45.4545 | 0.0000 | 0.0000 | 0.0000 | 9.0909 | 269 |
| 6 | 0.0000 | 22.2222 | 22.2222 | 44.4444 | 0.0000 | 0.0000 | 0.0000 | 11.1111 | 301 |
| 7 | 20.0000 | 20.0000 | 20.0000 | 40.0000 | 0.0000 | 0.0000 | 0.0000 | 0.0000 | 116 |
| 8 | 0.0000 | 25.0000 | 25.0000 | 25.0000 | 0.0000 | 25.0000 | 0.0000 | 0.0000 | 472 |
| 9 | 25.0000 | 25.0000 | 25.0000 | 25.0000 | 0.0000 | 0.0000 | 0.0000 | 0.0000 | 135 |
| 10 | 25.0000 | 25.0000 | 25.0000 | 25.0000 | 0.0000 | 0.0000 | 0.0000 | 0.0000 | 135 |
| 11 | 20.0000 | 20.0000 | 20.0000 | 0.0000 | 0.0000 | 40.0000 | 0.0000 | 0.0000 | 509 |
| 12 | 0.0000 | 19.2308 | 19.2308 | 38.4615 | 0.0000 | 3.8462 | 19.2308 | 0.0000 | 161 |
| 13 | 0.0000 | 18.5185 | 18.5185 | 37.0370 | 0.0000 | 7.4074 | 18.5185 | 0.0000 | 200 |
| 14 | 0.0000 | 18.1818 | 18.1818 | 36.3636 | 0.0000 | 9.0909 | 18.1818 | 0.0000 | 235 |
| 15 | 0.0000 | 18.1818 | 18.1818 | 36.3636 | 0.0000 | 9.0909 | 18.1818 | 0.0000 | 218 |
| 16 | 0.0000 | 17.5439 | 17.5439 | 35.0877 | 0.0000 | 12.2807 | 17.5439 | 0.0000 | 275 |
| 17 | 20.0000 | 20.0000 | 20.0000 | 35.0000 | 0.0000 | 5.0000 | 0.0000 | 0.0000 | 110 |
| 18 | 0.0000 | 17.2414 | 17.2414 | 34.4828 | 0.0000 | 13.7931 | 17.2414 | 0.0000 | 315 |
| 19 | 0.0000 | 16.6667 | 16.6667 | 33.3333 | 0.0000 | 16.6667 | 16.6667 | 0.0000 | 390 |
| 20 | 20.0000 | 20.0000 | 20.0000 | 32.5000 | 0.0000 | 7.5000 | 0.0000 | 0.0000 | 131 |
| 21 | 0.0000 | 16.1290 | 16.1290 | 32.2581 | 0.0000 | 19.3548 | 16.1290 | 0.0000 | 520 |
| 22 | 0.0000 | 15.3846 | 15.3846 | 30.7692 | 0.0000 | 23.0769 | 15.3846 | 0.0000 | 550 |
| 23 | 20.0000 | 20.0000 | 20.0000 | 30.0000 | 0.0000 | 10.0000 | 0.0000 | 0.0000 | 159 |
| 24 | 0.0000 | 14.7059 | 14.7059 | 29.4118 | 0.0000 | 26.4706 | 14.7059 | 0.0000 | 555 |
| 25 | 0.0000 | 14.2857 | 14.2857 | 28.5714 | 0.0000 | 28.5714 | 14.2857 | 0.0000 | 565 |
| 26 | 0.0000 | 13.8889 | 13.8889 | 27.7778 | 0.0000 | 30.5556 | 13.8889 | 0.0000 | 575 |
| 27 | 0.0000 | 13.3333 | 13.3333 | 26.6667 | 0.0000 | 33.3333 | 13.3333 | 0.0000 | 596 |
| 28 | 21.0526 | 21.0526 | 26.3158 | 26.3158 | 0.0000 | 5.2632 | 0.0000 | 0.0000 | 161 |
| 29 | 20.0000 | 20.0000 | 20.0000 | 25.0000 | 0.0000 | 15.0000 | 0.0000 | 0.0000 | 388 |
| 30 | 0.0000 | 23.8095 | 23.8095 | 23.8095 | 0.0000 | 23.8095 | 0.0000 | 4.7619 | 549 |
| 31 | 23.2558 | 23.2558 | 23.2558 | 23.2558 | 0.0000 | 0.0000 | 0.0000 | 6.9767 | 200 |
| 32 | 20.0000 | 20.0000 | 20.0000 | 22.5000 | 0.0000 | 17.5000 | 0.0000 | 0.0000 | 538 |
| 33 | 22.2222 | 22.2222 | 0.0000 | 22.2222 | 0.0000 | 22.2222 | 11.1111 | 0.0000 | 496 |
| 34 | 22.2222 | 22.2222 | 22.2222 | 22.2222 | 0.0000 | 11.1111 | 0.0000 | 0.0000 | 247 |
| 35 | 22.2222 | 0.0000 | 22.2222 | 22.2222 | 0.0000 | 22.2222 | 0.0000 | 11.1111 | 605 |
| 36 | 0.0000 | 22.2222 | 22.2222 | 22.2222 | 0.0000 | 22.2222 | 0.0000 | 11.1111 | 622 |
| 37 | 22.2222 | 22.2222 | 22.2222 | 22.2222 | 0.0000 | 0.0000 | 0.0000 | 11.1111 | 310 |
| 38 | 22.2222 | 22.2222 | 22.2222 | 22.2222 | 0.0000 | 0.0000 | 0.0000 | 11.1111 | 220 |
| 39 | 0.0000 | 20.8333 | 20.8333 | 20.8333 | 0.0000 | 20.8333 | 0.0000 | 16.6667 | 854 |
| 40 | 20.6186 | 20.6186 | 20.6186 | 20.6186 | 0.0000 | 0.0000 | 0.0000 | 17.5258 | 420 |
| 41 | 20.0000 | 20.0000 | 0.0000 | 20.0000 | 0.0000 | 20.0000 | 20.0000 | 0.0000 | 419 |
| 42 | 20.0000 | 20.0000 | 20.0000 | 20.0000 | 0.0000 | 20.0000 | 0.0000 | 0.0000 | 520 |
| 43 | 20.0000 | 20.0000 | 20.0000 | 20.0000 | 0.0000 | 20.0000 | 0.0000 | 0.0000 | 484 |
| 44 | 20.0000 | 20.0000 | 20.0000 | 20.0000 | 0.0000 | 20.0000 | 0.0000 | 0.0000 | 395 |
| 45 | 0.0000 | 20.0000 | 20.0000 | 20.0000 | 0.0000 | 20.0000 | 20.0000 | 0.0000 | 342 |
| 46 | 0.0000 | 20.0000 | 20.0000 | 20.0000 | 0.0000 | 20.0000 | 0.0000 | 20.0000 | 912 |
| 47 | 0.0000 | 20.0000 | 20.0000 | 20.0000 | 20.0000 | 0.0000 | 20.0000 | 0.0000 | 296 |
| 48 | 20.0000 | 20.0000 | 20.0000 | 20.0000 | 20.0000 | 0.0000 | 20.0000 | 0.0000 | 133 |
| 49 | 20.0000 | 20.0000 | 20.0000 | 20.0000 | 0.0000 | 0.0000 | 20.0000 | 0.0000 | 286 |
| 50 | 20.0000 | 20.0000 | 20.0000 | 20.0000 | 20.0000 | 0.0000 | 0.0000 | 0.0000 | 144 |
| 51 | 20.0000 | 0.0000 | 20.0000 | 20.0000 | 20.0000 | 0.0000 | 20.0000 | 0.0000 | 209 |
| 52 | 0.0000 | 20.0000 | 20.0000 | 20.0000 | 0.0000 | 0.0000 | 20.0000 | 20.0000 | 263 |
| 53 | 18.1818 | 18.1818 | 18.1818 | 18.1818 | 0.0000 | 27.2727 | 0.0000 | 0.0000 | 402 |
| 54 | 16.6667 | 16.6667 | 16.6667 | 16.6667 | 0.0000 | 33.3333 | 0.0000 | 0.0000 | 432 |
| 55 | 15.3846 | 15.3846 | 15.3846 | 15.3846 | 0.0000 | 38.4615 | 0.0000 | 0.0000 | 487 |
| 56 | 20.0000 | 20.0000 | 20.0000 | 15.0000 | 0.0000 | 25.0000 | 0.0000 | 0.0000 | 487 |
| 57 | 14.2857 | 14.2857 | 14.2857 | 14.2857 | 0.0000 | 42.8571 | 0.0000 | 0.0000 | 506 |
| 58 | 0.0000 | 23.2558 | 34.8837 | 11.6279 | 23.2558 | 6.9767 | 0.0000 | 0.0000 | 297 |
| 59 | 0.0000 | 22.2222 | 33.3333 | 11.1111 | 22.2222 | 11.1111 | 0.0000 | 0.0000 | 396 |
| 60 | 20.0000 | 20.0000 | 20.0000 | 10.0000 | 0.0000 | 30.0000 | 0.0000 | 0.0000 | 484 |
| 61 | 22.2222 | 22.2222 | 22.2222 | 0.0000 | 0.0000 | 22.2222 | 0.0000 | 11.1111 | 904 |
| 62 | 15.3846 | 15.3846 | 15.3846 | 15.3846 | 0.0000 | 30.7692 | 0.0000 | 7.6923 | 615 |
| 63 | 16.6667 | 16.6667 | 16.6667 | 16.6667 | 0.0000 | 25.0000 | 0.0000 | 8.3333 | 665 |
| 64 | 10.0000 | 20.0000 | 20.0000 | 20.0000 | 0.0000 | 20.0000 | 0.0000 | 10.0000 | 780 |
| 65 | 19.6078 | 19.6078 | 11.7647 | 19.6078 | 0.0000 | 19.6078 | 0.0000 | 9.8039 | 750 |
| 66 | 14.2857 | 14.2857 | 14.2857 | 14.2857 | 0.0000 | 28.5714 | 14.2857 | 0.0000 | 560 |
| 67 | 15.3846 | 30.7692 | 15.3846 | 15.3846 | 0.0000 | 15.3846 | 0.0000 | 7.6923 | 864 |
| 68 | 16.6667 | 25.0000 | 16.6667 | 16.6667 | 0.0000 | 16.6667 | 0.0000 | 8.3333 | 815 |
| 69 | 18.1818 | 18.1818 | 18.1818 | 18.1818 | 0.0000 | 18.1818 | 0.0000 | 9.0909 | 770 |
| 70 | 18.1818 | 18.1818 | 18.1818 | 18.1818 | 0.0000 | 18.1818 | 0.0000 | 9.0909 | 730 |
| 71 | 18.1818 | 18.1818 | 18.1818 | 18.1818 | 0.0000 | 18.1818 | 0.0000 | 9.0909 | 725 |
| 72 | 18.1818 | 18.1818 | 18.1818 | 18.1818 | 0.0000 | 18.1818 | 0.0000 | 9.0909 | 720 |
| 73 | 20.0000 | 10.0000 | 20.0000 | 20.0000 | 0.0000 | 20.0000 | 0.0000 | 10.0000 | 612 |
| 74 | 16.6667 | 16.6667 | 25.0000 | 16.6667 | 0.0000 | 16.6667 | 0.0000 | 8.3333 | 640 |
| 75 | 25.0000 | 16.6667 | 16.6667 | 16.6667 | 0.0000 | 16.6667 | 0.0000 | 8.3333 | 711 |
| 76 | 16.6667 | 16.6667 | 16.6667 | 25.0000 | 0.0000 | 16.6667 | 0.0000 | 8.3333 | 627 |
| 77 | 15.3846 | 15.3846 | 30.7692 | 15.3846 | 0.0000 | 15.3846 | 0.0000 | 7.6923 | 650 |
| 78 | 30.7692 | 15.3846 | 15.3846 | 15.3846 | 0.0000 | 15.3846 | 0.0000 | 7.6923 | 596 |
| 79 | 15.3846 | 15.3846 | 15.3846 | 30.7692 | 0.0000 | 15.3846 | 0.0000 | 7.6923 | 404 |
| 80 | 20.0000 | 20.0000 | 20.0000 | 20.0000 | 0.0000 | 10.0000 | 0.0000 | 10.0000 | 440 |
| 81 | 18.1818 | 18.1818 | 18.1818 | 18.1818 | 0.0000 | 18.1818 | 9.0909 | 0.0000 | 458 |
| 82 | 18.1818 | 18.1818 | 18.1818 | 9.0909 | 0.0000 | 18.1818 | 18.1818 | 0.0000 | 423 |
| 83 | 9.0909 | 18.1818 | 18.1818 | 18.1818 | 0.0000 | 18.1818 | 18.1818 | 0.0000 | 473 |
| 84 | 18.1818 | 18.1818 | 9.0909 | 18.1818 | 0.0000 | 18.1818 | 18.1818 | 0.0000 | 418 |
| 85 | 16.6667 | 16.6667 | 16.6667 | 16.6667 | 0.0000 | 16.6667 | 16.6667 | 0.0000 | 416 |
| 86 | 16.6667 | 16.6667 | 16.6667 | 16.6667 | 0.0000 | 16.6667 | 16.6667 | 0.0000 | 420 |
| 87 | 16.6667 | 16.6667 | 16.6667 | 16.6667 | 0.0000 | 16.6667 | 16.6667 | 0.0000 | 350 |
| 88 | 18.1818 | 9.0909 | 18.1818 | 18.1818 | 0.0000 | 18.1818 | 18.1818 | 0.0000 | 367 |
| 89 | 18.1818 | 18.1818 | 18.1818 | 18.1818 | 0.0000 | 9.0909 | 18.1818 | 0.0000 | 208 |
| 90 | 18.1818 | 18.1818 | 18.1818 | 18.1818 | 0.0000 | 9.0909 | 18.1818 | 0.0000 | 208 |
| 91 | 18.1818 | 18.1818 | 18.1818 | 18.1818 | 0.0000 | 9.0909 | 18.1818 | 0.0000 | 225 |

| 92 | 18.8679 | 18.8679 | 18.8679 | 18.8679 | 18.8679 | 5.6604 | 0.0000 | 0.0000 | 125 |
| 93 | 18.0000 | 35.0000 | 10.0000 | 5.5000 | 7.5000 | 24.0000 | 0.0000 | 0.0000 | 650 |
| 94 | 9.0000 | 35.0000 | 10.0000 | 5.0000 | 15.5000 | 25.5000 | 0.0000 | 0.0000 | 628 |
| 95 | 6.0000 | 35.0000 | 6.0000 | 5.0000 | 18.0000 | 30.0000 | 0.0000 | 0.0000 | 605 |
| 96 | 18.0000 | 22.0000 | 22.0000 | 22.0000 | 5.0000 | 11.0000 | 0.0000 | 0.0000 | 198 |
| 97 | 16.0000 | 18.5000 | 16.5000 | 13.5000 | 5.0000 | 30.5000 | 0.0000 | 0.0000 | 505 |

Table S-12: Test Results in Reference (61)

**Optimal hardness in the paper (61)**

| Elements | Co (at%) | Cr (at%) | Fe (at%) | Ni (at%) | Mn (at%) | Al (at%) | Cu (at%) | Mo (at%) | Optimal Hardness |
|---|---|---|---|---|---|---|---|---|---|
| Percentage | 18 | 35 | 10 | 5.5 | 7.5 | 24 | 0 | 0 | 650 |

Table: S-13: Optimal hardness in the paper (57)

**Meta Analysis**

**CSR Equation**

Hardness = $(1.04 \times 10^{-3})$

+ $(1.92 \times 10^{-2})$ $c_1$ + $(-1.03 \times 10^{-1})$ $c_1^2$ + $(3.35 \times 10^{-2})$ $c_2$ + $(3.96 \times 10^{-1})$ $c_2^2$ + $(-3.94 \times 10^{-3})$ $c_3$ + $(5.35 \times 10^{-1})$ $c_3^2$

+ $(-3.4 \times 10^{-2})$ $c_4$ + $(-6.44 \times 10^{-2})$ $c_4^2$ + $(1.96 \times 10^{-2})$ $c_5$ + $(4.71 \times 10^{-1})$ $c_5^2$ + $(2.08 \times 10^{-2})$ $c_6$ + $(-5.22 \times 10^{-2})$ $c_6^2$

+ $(2.6 \times 10^{-2})$ $c_7$ + $(-4.23 \times 10^{-1})$ $c_7^2$ + $(2.27 \times 10^{-2})$ $c_8$ + $(1.27 \times 10^{-1})$ $c_8^2$ + $(3.23 \times 10^{-1})$ $c_7 c_8$ + $(1.02)$ $c_6 c_8$

+ $(2.1 \times 10^{-1})$ $c_6 c_7$ + $(5.56 \times 10^{-15})$ $c_5 c_8$ + $(1.94)$ $c_5 c_7$ + $(-1.16 \times 10^{-2})$ $c_5 c_6$ + $(-2.02 \times 10^{-1})$ $c_4 c_8$ + $(-3.44 \times 10^{-1})$ $c_4 c_7$

+ $(3.3 \times 10^{-1})$ $c_4 c_6$ + $(-3.09)$ $c_4 c_5$ + $(-1.54)$ $c_3 c_8$ + $(4.85 \times 10^{-2})$ $c_3 c_7$ + $(1.22)$ $c_3 c_6$ + $(8.2 \times 10^{-1})$ $c_3 c_5$

+ $(-1.36 \times 10^{-1})$ $c_3 c_4$ + $(2.17)$ $c_2 c_8$ + $(7.02 \times 10^{-1})$ $c_2 c_7$ + $(-6.02 \times 10^{-1})$ $c_2 c_6$ + $(8.05 \times 10^{-1})$ $c_2 c_5$ + $(9.16 \times 10^{-1})$ $c_2 c_4$

+ $(-1.85)$ $c_2 c_3$ + $(3.82 \times 10^{-1})$ $c_1 c_8$ + $(1.43 \times 10^{-1})$ $c_1 c_7$ + $(-3.88 \times 10^{-2})$ $c_1 c_6$ + $(1.02)$ $c_1 c_5$ + $(-8.07 \times 10^{-1})$ $c_1 c_4$ + $(5.07 \times 10^{-1})$ $c_1 c_3$ + $(8.14 \times 10^{-1})$ $c_1 c_2$

$R^2 = 0.93$

| Hardness (Hv) | Optimal hardness identified by ANN | Optimal hardness identified by CSR | Optimized hardness identified by experiment |
|---|---|---|---|
| Unit: Hv | 670 | 642 | 650 |

Table S-14: Comparison of optimal hardness identified by ANN, CSR and Experiment

**Conclusion:**

The optimal hardness predicted by the CSR equation is consistent with both the ANN predictions and the experimental results.

# 7 Data for the Social Complex Systems

**Data Sources:**

City Lives (62): http://www.worldcitiescultureforum.com/data

GDP (63): World Bank, US Bureau of Economic Analysis, International Monetary Fund, Statistics Canada

**Data for the CSR analysis of social complex systems**

| City | Population | | Area | | 1/Area | | Art Galleries | | Book Store | | Restaurant | | GDP | | | GDP/Population | |
|---|---|---|---|---|---|---|---|---|---|---|---|---|---|---|---|---|
| | P | LOG P | A km^2 | LOG A | 1/A | LOG (1/A) | AG | LOG AG | BS | LOG BS | R | LOG R | G $billion | LOG G | G/P | LOG(G/P) |
| Amsterdam | 873338 | 5.94 | 2580 | 3.41 | 3.90E-04 | -3.41 | 33 | 1.52 | 110 | 2.04 | 2293 | 3.36 | 222 | 11.35 | 254197.1 | 5.41 |
| Austin | 1026833 | 6.01 | 845 | 2.93 | 1.18E-03 | -2.93 | 32 | 1.51 | 30 | 1.48 | 2104 | 3.32 | 159 | 11.2 | 216198.7 | 5.33 |
| Barcelona | 1636762 | 6.21 | 101.4 | 2.01 | 9.86E-03 | -2.01 | 454 | 2.66 | 324 | 2.51 | 2916 | 3.46 | 250 | 11.4 | 135633.7 | 5.13 |
| Dublin | 1347359 | 6.13 | 918 | 2.96 | 1.09E-03 | -2.96 | 49 | 1.69 | 67 | 1.83 | 431 | 2.63 | 162 | 11.21 | 164766.8 | 5.22 |
| Edinburgh | 527620 | 5.72 | 263 | 2.42 | 3.80E-03 | -2.42 | 27 | 1.43 | 20 | 1.3 | 960 | 2.98 | 73 | 10.86 | 420757.4 | 5.62 |
| Helsinki | 653835 | 5.82 | 719 | 2.86 | 1.39E-03 | -2.86 | 100 | 2 | 20 | 1.3 | 150 | 2.18 | 91 | 10.96 | 339535.2 | 5.53 |
| Hong Kong | 7428300 | 6.87 | 2755 | 3.44 | 3.60E-04 | -3.44 | 126 | 2.1 | 1070 | 3.03 | 15863 | 4.2 | 366 | 11.56 | 29885.71 | 4.48 |
| Istanbul | 15840900 | 7.2 | 5461 | 3.74 | 1.80E-04 | -3.74 | 98 | 1.99 | 670 | 2.83 | 23238 | 4.37 | 233 | 11.37 | 14014.36 | 4.15 |
| lisbon | 544851 | 5.74 | 100 | 2 | 1.00E-02 | -2 | 78 | 1.89 | 196 | 2.29 | 4385 | 3.64 | 94 | 10.97 | 407450.8 | 5.61 |
| London | 9002488 | 6.95 | 1572 | 3.2 | 6.40E-04 | -3.2 | 478 | 2.68 | 219 | 2.34 | 14745 | 4.17 | 565 | 11.75 | 24659.85 | 4.39 |
| Los Angeles | 10004019 | 7 | 10510 | 4.02 | 1.00E-04 | -4.02 | 230 | 2.36 | 177 | 2.25 | 30800 | 4.49 | 1044 | 12.02 | 22191.08 | 4.35 |
| Melbourne | 5159211 | 6.71 | 9992 | 4 | 1.00E-04 | -4 | 162 | 2.21 | 1649 | 3.22 | 4620 | 3.66 | 176 | 11.25 | 43029.84 | 4.63 |
| Milan | 1404431 | 6.15 | 182 | 2.26 | 5.49E-03 | -2.26 | 192 | 2.28 | 269 | 2.43 | 5856 | 3.77 | 315 | 11.5 | 158071.1 | 5.2 |
| Montreal | 2004265 | 6.3 | 499 | 2.7 | 2.00E-03 | -2.7 | 72 | 1.86 | 108 | 2.03 | 5296 | 3.72 | 151 | 11.18 | 110763.8 | 5.04 |
| lan Francisc | 844363 | 5.93 | 121 | 2.08 | 8.26E-03 | -2.08 | 98 | 1.99 | 77 | 1.89 | 4045 | 3.61 | 501 | 11.7 | 262920.1 | 5.42 |
| Seoul | 10010983 | 7 | 605.24 | 2.78 | 1.65E-03 | -2.78 | 475 | 2.68 | 530 | 2.72 | 125740 | 5.1 | 779 | 11.89 | 22175.64 | 4.35 |
| Shenzhen | 11908400 | 7.08 | 1997 | 3.3 | 5.00E-04 | -3.3 | 400 | 2.6 | 709 | 2.85 | | | 432 | 11.64 | 18642.3 | 4.27 |
| Stockholm | 977345 | 5.99 | 16541 | 4.22 | 6.00E-05 | -4.22 | 152 | 2.18 | 217 | 2.34 | 5814 | 3.76 | 152 | 11.18 | 227146 | 5.36 |
| Sydney | 4822991 | 6.68 | 12368 | 4.09 | 8.00E-05 | -4.09 | 170 | 2.23 | 258 | 2.41 | 19384 | 4.29 | 356 | 11.55 | 46019.99 | 4.66 |
| Tokyo | 13843525 | 7.14 | 2194 | 3.34 | 4.60E-04 | -3.34 | 416 | 2.62 | 1579 | 3.2 | 137669 | 5.14 | 1688 | 12.23 | 16036.38 | 4.21 |
| Toronto | 6346088 | 6.8 | 630 | 2.8 | 1.59E-03 | -2.8 | 433 | 2.64 | 365 | 2.56 | 7984 | 3.9 | 332 | 11.52 | 34962.18 | 4.54 |
| Warsaw | 1794200 | 6.25 | 517 | 2.71 | 1.93E-03 | -2.71 | 100 | 2 | 195 | 2.29 | 4808 | 3.68 | 103 | 11.01 | 123732 | 5.09 |

Table S-15: Data for the CSR analysis of social complex system

# References

1.   A. F. Siegenfeld and Y. Bar-Yam, An Introduction to Complex Systems Science and Its Applications, *Complexity,* 2020, Article ID 6105872, https://doi.org/10.1155/2020/6105872, (2020)

2.   Y. Holovatch, R. Kenna, S. Thurner, Complex systems: physics beyond physics. *Eur. J. Phys.* 38, 23002 (2017), doi:10.1088/1361-6404/aa5a87.

3.   J. M. Ottino, Complex Systems. *AIChE Journal.* 49, 292–299 (2003), doi:10.1002/aic.690490202.

4.   H. A. Simon, *"The Architecture of Complexity."* Proceedings of the American Philosophical Society 106 (6): 467-482. (1962)

5.   U. Alon, Network motifs: Theory and experimental approaches. *Nature Reviews Genetics*, 8(6), 450–461 (2007). https://doi.org/10.1038/nrg2102

6.   A.-L, Barabási, & Z.N. Oltvai, Network biology: Understanding the cell's functional organization. *Nature Reviews Genetics*, 5(2), 101–113. (2004). https://doi.org/10.1038/nrg1272

7.   S., et al. Camazine, *Self-Organization in Biological Systems*. Princeton University Press. (2003). ISBN: 978-0691116242

8.   M. C. Cross, & P. C. Hohenberg, Pattern formation outside of equilibrium. *Reviews of Modern Physics*, 65(3), 851–1112. (1993). https://doi.org/10.1103/RevModPhys.65.851

9.   P. Bak, C., Tang, &  K. Wiesenfeld, Self-organized criticality: An explanation of 1/f noise. *Physical Review Letters*, 59(4), 381–384. (1987). https://doi.org/10.1103/PhysRevLett.59.381

10. T. Mora, & W. Bialek, (2011). Are biological systems poised at criticality? *Journal of Statistical Physics*, 144(2), 268–302. https://doi.org/10.1007/s10955-011-0229-4

11. E. Ising, Beitrag zur Theorie des Ferromagnetismus. *Z. Physik*. 31, 253–258 (1925), doi:10.1007/BF02980577.

12. C. N. Yang, The Spontaneous Magnetization of a Two-Dimensional Ising Model. *Physical Review.* 85, 808–816 (1952), doi:10.1103/PhysRev.85.808.

13. G. Parisi, Number of Order Parameters for Spin-Glasses, *Physical Review Letters.* 43, 1754- 1756(1979), doi:10.1103/PhysRevLett.43.1754

14. R. A. Fisher, The Wave of Advance of Advantageous Genes, *Annals of Eugenics*, https://doi.org/10.1111/j.1469-1809.1937.tb02153.x, 1937

15. F. Black, & M. Scholes, (1973). "The Pricing of Options and Corporate Liabilities." *Journal of Political Economy*, 81(3), 637–654.

16. G. E. P. Box, K. B. Wilson, "On the Experimental Attainment of Optimum Conditions", Journal of the Royal Statistical Society: Series B, (Methodological), 13 (1) 1-38 (1951), doi.org/10.1111/j.2517-6161.1951.tb00067.x.

17. J. Pearl, Causality: Models, reasoning, and inference, Reviewed by Leland Gerson Neuberg, Econometric Theory (Cambridge University Press, Cambridge U.K., New York, 2003).

18. Myers, R. H., Montgomery, D. C., & Anderson-Cook, C. M. (2016). *Response Surface Methodology: Process and Product Optimization Using Designed Experiments* (4th ed.).

19. Wiley. 19. Khuri, A. I., & Mukhopadhyay, S. (2010). Response surface methodology. *Wiley Interdisciplinary Reviews: Computational Statistics*, 2(2), 128–149. https://doi.org/10.1002/wics.73

20. N. R. Draper and D. K. J. Lin, (1996). *Response Surface Designs*. In S. Ghosh & C. R. Rao (Eds.), Handbook of Statistics, Vol. 13 (pp. 343–390). Elsevier.

21. R. Gallagher, T. Appenzeller, Beyond Reductionism. *Science*. 284, 79 (1999), doi:10.1126/science.284.5411.79.

22. P. W. Anderson, More is Different, Science, 177(4047), 393–396 (1972).

23. J. Horgan, From Complexity to Perplexity, Scientific American v. 272(6), 104–109 (1995), DOI: 10.1038/scientificamerican0695-104

24. I. Newton, *Philosophiæ Naturalis Principia Mathematica*. London: Joseph Streater for the Royal Society, (1687)

25. I. B. Cohen and A. Whitman Translated *The Principia: Mathematical Principles of Natural Philosophy*. University of California Press, (1999).

26. L. Euler, . *Mechanica sive motus scientia analytice exposita,* Volumen I: De motu rectilineo. Academiae Scientiarum Imperialis Petropolitanae. (1736).

27. I. Al-Shyoukh, F. Yu, J. Feng, K. Yan, S. Dubinett, C.M. Ho, J.S. Shamma, R. Sun, Systematic quantitative characterization of cellular responses induced by multiple signals. *BMC systems biology*. 5, 88 (2011), doi:10.1186/1752-0509-5-88.

28. P. K. Wong, F. Yu, A. Shahangian, G. Cheng, R. Sun, C. M. Ho, Closed-loop control of cellular functions using combinatory drugs guided by a stochastic search algorithm. *Proceedings of the National Academy of Sciences of the United States of America*. 105, 5105–5110 (2008), doi:10.1073/pnas.0800823105.

29. D. Ho, C. M. Ho, System control-mediated drug delivery towards complex systems via nanodiamond carriers. *International Journal of Smart and Nano Materials*. 1, 69–81 (2010), doi:10.1080/19475411003619736.

30. H. Tsutsui, B. Valamehr, A. Hindoyan, R. Qiao, X. Ding, S. Guo, O. N. Witte, X. Liu, C. M. Ho, H. Wu, An optimized small molecule inhibitor cocktail supports long-term maintenance of human embryonic stem cells. *Nat Commun*. 2, 167 (2011), doi:10.1038/ncomms1165.

31. A. Weiss, R. H. Berndsen, X. Ding, C. M. Ho, P. J. Dyson, H. van den Bergh, A. W. Griffioen, P. Nowak-Sliwinska, A streamlined search technology for identification of synergistic drug combinations. *Scientific reports*. 5, 14508 (2015), doi:10.1038/srep14508.

32. Patrycja Nowak-Sliwinska1, Andrea Weiss1, Xianting Ding, Paul J Dyson, Hubert van den Bergh, Arjan W Griffioen & Chih-Ming Ho, "Optimization of drug combinations using Feedback System Control", Nature Protocols, VOL.11 NO.2, pp.302-315, 2016

33. A. Silva, B. Y. Lee, D. L. Clemens, T. Kee, X. Ding, C. M. Ho, M. A. Horwitz, Output-driven feedback system control platform optimizes combinatorial therapy of tuberculosis using a macrophage cell culture model. *Proceedings of the National Academy of Sciences of the United States of America*. 113, E2172-9 (2016), doi:10.1073/pnas.1600812113.

34. X. Ding, W. Liu, A. Weiss, Y. Li, I. Wong, A. W. Griffioen, H. van den Bergh, H. Xu, P. Nowak-Sliwinska, C. M. Ho, Discovery of a low order drug-cell response surface for applications in personalized medicine. *Physical biology*. 11, 65003 (2014), doi:10.1088/1478-3975/11/6/065003.

35. M. B. M. A. Rashid, T. B. Toh, L. Hooi, A. Silva, Y. Zhang, P. F. Tan, A. L. Teh, N. Karnani, S. Jha, C. M. Ho, W. J. Chng, D. Ho, E. K. H. Chow, Optimizing drug combinations against multiple myeloma using a quadratic phenotypic optimization platform (QPOP). *Sci. Transl. Med.* 10 (2018), doi:10.1126/scitranslmed.aan0941.

36. B. Y. Lee, D. L. Clemens, A. Silva, B. J. Dillon, S. Masleša-Galić, S. Nava, X. Ding, C. M. Ho, M. A. Horwitz, Drug regimens identified and optimized by output-driven platform markedly reduce tuberculosis treatment time. *Nat. Commun.* 8, 14183 (2017), doi:10.1038/ncomms14183.

37. X. Ding, V. H. S. Chang, Y. Li, X. Li, H. Xu, C. M. Ho, D. Ho, Y. Yen, Harnessing an Artificial Intelligence Platform to Dynamically Individualize Combination Therapy for Treating Colorectal Carcinoma in a Rat Model. *Adv. Therap.* 3, 1900127 (2020), doi:10.1002/adtp.201900127.

38. A. Zarrinpar, D. K. Lee, A. Silva, N. Datta, T. Kee, C. Eriksen, K. Weigle, V.Agopian, F. Kaldas, D. Farmer, S. E. Wang, R. Busuttil, C. M. Ho, D. Ho, Individualizing liver transplant immunosuppression using a phenotypic personalized medicine platform. *Sci. Transl. Med.* 8, 333ra49 (2016), doi:10.1126/scitranslmed.aac5954.

39. A. J. Pantuck, D. K. Lee, T. Kee, P. Wang, S. Lakhotia, M. H. Silverman, C. Mathis, A. Drakaki, A. S. Belldegrun, C. M. Ho, D. Ho, Modulating BET Bromodomain Inhibitor ZEN-3694 and Enzalutamide Combination Dosing in a Metastatic Prostate Cancer Patient Using CURATE.AI, an Artificial Intelligence Platform. *Adv. Therap.* 1, 1800104 (2018), doi:10.1002/adtp.201800104.

40. T. Kee, W. Chee; A. Blasiak, P. Wang, J. K. Chong, J. Chen, B. T. T. Yeo, D. Ho, C. L. Asplund, Harnessing CURATE.AI as a Digital Therapeutics Platform by Identifying Identifying N-of-1 Learning Trajectory Profiles. *Adv. Therap.* 2 (2019), doi:10.1002/adtp.201900023.

41. Y. Shen, T. Liu, J. Chen, X. Li, L. Liu, J. Shen, J. Wang, R. Zhang, M. Sun, Z. Wang, W.Song, T. Qi, Y. Tang, X. Meng, L. Zhang, D. Ho, Dean; C. M. Ho, X. Ding, H. Z. Lu, Harnessing Artificial Intelligence to Optimize Long-Term Maintenance Dosing for Antiretroviral-Naive Adults with HIV-1 Infection. *Adv. Therap.* 3 (2020), doi:10.1002/adtp.201900114.

42. X. Zheng, X. Gui, L. Yao, J. Ma, Y. He, H. Lou, J. Gu, R. Ying, L. Chen, Q. Sun, Y. Liu, C. -M. Ho, B. -Y. Lee, D. L. Clemens, M. A. Horwitz, X. Ding, X. Hao, H. Yang and W. Sha, "Efficacy and safety of an innovative short-course regimen containing clofazimine for treatment of drug-susceptible tuberculosis: a clinical trial", (2023), Emerging Microbes & Infections, DOI: 10.1080/22221751.2023.2187247.

43. J. Khong , M. Lee, C. Warren, U.B. Kim, S. Duarte, K. A. Andreoni, S. Shrestha3, M. W. Johnson, N. R. Battula, D. M.McKimmy, T. Beduschi, J.-H Lee, D. M. Li, C. -M. Ho & A. Zarrinpar , "Tacrolimus dosing in liver transplant recipients using phenotypic personalized medicine: A phase 2 randomized clinical trial", (2025) Nature Communications, https://doi.org/10.1038/s41467-025-59739-6

44. A. Blasiak, J. J. Lim, S Seah, T. Kee, A. Remus, D. H. Chye, P. S. Wong, L. Hooi, A. T. Truong, N, Le1, C. Chan, R. Desai, X. Ding, B. Hanson, E. Chow, D. Ho, "IDentif.AI: Rapidly optimizing combination therapy design against severe Acute Respiratory Syndrome Coronavirus 2 (SARS-Cov-2) with digital drug development", Bioeng Transl Med. (2021;6:e10196), https://doi.org/10.1002/btm2.10196

45. X.T. Ding, H.Q. Xu, C. Hopper , J. Yang , C.M. Ho, Use of Fractional Factorial Designs in Antiviral Drug Studies", Quality and Reliability Engineering International, DOI: 10.1002/qre.1308, 2012

46. J. Luna, J. Jaynes, H. Q. Xu, W. K. Wong, "Orthogonal array composite designs for drug combination experiments with applications for tuberculosis",Stat Med. Jul 30;41(17):3380-3397. (2022), DOI: 10.1002/sim.94, 2022

45. E. B. Maldonado, S. Parsons, E. Y. Chen, A. Haslam, V. Prasad, Estimation of US patients with cancer who may respond to cytotoxic chemotherapy. *Future science OA*. 6, FSO600 (2020), doi:10.2144/fsoa-2020-0024.

46. K. D. Miller, L. Nogueira, T. Devasia, A. B. Mariotto, K. R. Yabroff, A. Jemal, J. Kramer, R. L. Siegel, Cancer treatment and survivorship statistics, 2022. *CA: a cancer journal for clinicians*. 72, 409–436 (2022), doi:10.3322/caac.21731.

47. D. Ho, S. R. Quake, E. R. B. McCabe, W. J. Chng, E. K. Chow, X. Ding, B. D. Gelb, G. S. Ginsburg, J. Hassenstab, C. M. Ho, W. C. Mobley, G. P. Nolan, S. T. Rosen, P. Tan, Y. Yen, A. Zarrinpar, Enabling Technologies for Personalized and Precision Medicine. *Trends in biotechnology*. 38, 497–518 (2020), doi:10.1016/j.tibtech.2019.12.021.

48. Kaplan, J., McCandlish, S., Henighan, T., et al. (2020). *Scaling Laws for Neural Language Models*. *arXiv preprint* arXiv:2001.08361. DOI: 10.48550/arXiv.2001.08361

49. E. M. Sefene, State-of-the-art of selective laser melting process: A comprehensive review. *Journal of Manufacturing Systems*. 63, 250–274 (2022), doi:10.1016/j.jmsy.2022.04.002.

50. Y. M. Wang, T. Voisin, J. T. McKeown, J. Ye, N. P. Calta, Z. Li, Z. Zeng, Y. Zhang, W. Chen, T. T. Roehling, R. T. Ott, M. K. Santala, P. J. Depond, M. J. Matthews, A. V. Hamza, T. Zhu, Additively manufactured hierarchical stainless steels with high strength and ductility. *Nat. Mater.* 17, 63–71 (2018), doi:10.1038/nmat5021.

51. Y. Zhang, G. S. Hong, D. Ye, K. Zhu, J. Y. Fuh, Extraction and evaluation of melt pool, plume and spatter information for powder-bed fusion AM process monitoring. *Materials & Design*. 156, 458–469 (2018), doi:10.1016/j.matdes.2018.07.002.

52. J. Kluczyński, L. Śnieżek, K. Grzelak, J. Janiszewski, P. Płatek, J. Torzewski, I. Szachogłuchowicz, K. Gocman, Influence of Selective Laser Melting Technological Parameters on the Mechanical Properties of Additively Manufactured Elements Using 316L Austenitic Steel. *Materials (Basel, Switzerland)*. 13 (2020), doi:10.3390/ma13061449.

53. J. Lei, Y. Ge, T. Liu, Z. Wei, Effects of Heat Treatment on the Microstructure and Mechanical Properties of Selective Laser Melting 316L Stainless Steel. *Shock and Vibration*. 2021, 1–12 (2021), doi:10.1155/2021/6547213.

54. B. Diepold, S. Neumeier, A. Meermeier, H.W. Höppel, T. Sebald, M. Göken, Temperature-Dependent Dynamic Strain Aging in Selective Laser Melted 316L. *Adv Eng Mater*. 23, 2001501 (2021), doi:10.1002/adem.202001501.

55. K. P. Lee, M. H. Lu, D, P. Tran, W. J. Chen, D, J.Yao, C. Chen, Optimization of tensile strength of nanotwinned Cu-Ni foils via complex system response methodology, Materials Science & Engineering, A 941, 148581, (2025), https://doi.org/10.1016/j.msea.2025.148581

56. Y. T. Fu, Y. H. Chang, C. M, Ho , and D. J. Yao, Optimization of tensile strength of nanotwinned Cu-Ni foils via complex system response methodology,(2024), DOI: 10.1109/impact63555.2024.10818925

57. Zalnezhad, E., Ahmed AD Sarhan, and M. Hamdi,  "Optimizing the PVD TiN thin film coating's parameters on aerospace AL7075-T6 alloy for higher coating hardness and adhesion with better tribological properties of the coating surface.", *The International Journal of Advanced Manufacturing Technology*, v. 64, 281-290, https://doi.org/10.1007/s00170-012-4022-6, 2013

58. Yu Zhou,  Xiaoyu Cui, Jianhua Weng, Saiyan Shi, Hua Han,  Chengmeng Chen, "Experimental investigation of the heat transfer performance of an oscillating heat pipe with graphene nanofluids", Powder Technology ,v. 332, 371–380, https://doi.org/10.1016/j.powtec.2018.02.048 , 2018

59. Chang, Y. J., Jui, C. Y., Lee, W. J., & Yeh, A. C. (2019). Prediction of the composition and hardness of high-entropy alloys by machine learning. *Jom*, *71*(10), 3433-3442., https://doi.org/10.1007/s11837-019-03704-4, 2019

60. City Lives: http://www.worldcitiescultureforum.com/data

61. GDP data: World Bank, US Bureau of Economic Analysis, Bureau of Economic Analysis, International Monetary Fund, Statistics Canada

62. L. M. A. Bettencourt and G. B. West, A unified theory of urban living, Nature, V. 467, 912-913 (2010) DOI: 10.1038/467912a

63. L. M. A. Bettencourt, J. Lobo, D. Helbing, C. Kuhnert, and G. B. West, "Growth, innovation, scaling, and the pace of life in cities", PNAS, v.104, 7301-7306 , (2007), www.pnas.org/cgi/content/full/0610172104/DC1 .

64. S. Pei, L. C. Parra, S. D. S. Reis, J. S. Andrade Jr., S. Havlin and H, A. Makse, Origins of power-law degree distribution in the heterogeneity of human activity in social networks, Scientific Reports, 3 : 1783, (2013)  DOI: 10.1038/srep01783

65. G. Boussinot, M. Apel, J.Zielinski, U, Hecht, and J.H. Schleifenbaum Strongly Out-of-Equilibrium Columnar Solidification During Laser Powder-Bed Fusion in Additive Manufacturing  Phys. Rev. Applied 11, 014025, 2019,

DOI:https://doi.org/10.1103/PhysRevApplied.11.014025

66. W. E. King, A. T. Anderson, R. M. Ferencz, N. E. Hodge, C. Kamath, S. A. Khairallah, Laser powder bed fusion additive manufacturing of metals: physics, computational, and materials challenges. *Applied Physics Reviews*. 2(4), 041304 (2015).

67. O. Shuleshova, W. Löser, D. Holland-Moritz, D. M. Herlach, J. Eckert, Solidification and melting of high temperature materials: in situ observations by synchrotron radiation. J Mater Sci. 47, 4497–4513 (2012), doi:10.1007/s10853-011-6184-2.

68. H. Tang, H. Huang, C. Liu, Z. Liu, W. Yan, Multi-Scale modelling of structure-property relationship in additively manufactured metallic materials. *International Journal of Mechanical Sciences*. 194, 106185 (2021).

69. C. Zhao, K. Fezzaa, R. W. Cunningham, H. Wen, F. Carlo, L. Chen, A. D. Rollett, T. Sun, Real-time monitoring of laser powder bed fusion process using high-speed X-ray imaging and diffraction. *Scientific reports*. 7, 3602 (2017), doi:10.1038/s41598-017-03761-2.

70. X.X Ding, P. Krutzik, A. A. Ghaffari, Y.  Zhaozhi, D. Miranda Jr., G. Cheng, C. -M. Ho, G. P. Nolan, D. J. Sanchez, "Cellular Signaling Analysis shows antiviral, ribavirin-mediated ribosomal signaling modulation", Antiviral Research 171:104598, (2019),  doi: 10.1016/j.antiviral.2019.104598